\documentclass[]{aa}
\usepackage{graphicx}
\usepackage{natbib}
\usepackage{amssymb,amsmath}
\usepackage{subfigure}
\usepackage[normalem]{ulem}
\bibpunct{(}{)}{;}{a}{,}{,}

\usepackage{pslatex}	    

\usepackage{ctable,rotating}
\newcommand{\p}{_{\text{p}}}

\newcommand{\sstar}{_{\star}}
\newcommand{\MM}{\mathcal{M}}
\newcommand{\E}{_{\oplus}}

\begin{document}

\title{Galactic cosmic rays on extrasolar Earth-like planets: II. Atmospheric implications}

\author{J.--M. Grie{\ss}meier
          \inst{1,2}
          \and
	  F. Tabataba-Vakili
	  \inst{3}
	  \and
          A. Stadelmann
          \inst{4}
	  \and
	  J. L. Grenfell 
	  \inst{5,6}
	  \and
	  D. Atri
	  \inst{7}
 }

\institute{
	LPC2E - Universit\'{e} d'Orl\'{e}ans / CNRS, 3A, Avenue de la Recherche Scientifique, 45071 Orléans cedex 2, France. \email{jean-mathias.griessmeier@cnrs-orleans.fr}
        \and
Station de Radioastronomie de Nan\c{c}ay, Observatoire de Paris - CNRS/INSU, USR 
  704 - Univ. Orl\'{e}ans, OSUC, route de Souesmes, 18330 Nan\c{c}ay, France
        \and
        Atmospheric, Oceanic and Planetary Physics, Department of Physics, University of Oxford, Clarendon Laboratory, Parks Road, Oxford OX1 3PU, UK
        \and
        Technische Universit\"{a}t Braunschweig, Mendelssohnstr. 3, 38106 Braunschweig, Germany
        \and
        Zentrum für Astronomie und Astrophysik (ZAA), Technische Universität Berlin (TUB), Hardenbergstr. 36, 10623 Berlin, Germany
        \and
        now at: Extrasolare Planeten und Atmosph\"{a}ren (EPA), Institut für Planetenforschung, Deutsches Zentrum für Luft- und Raumfahrt (DLR),
	Rutherfordstr. 2, 12489 Berlin, Germany
        \and
	Blue Marble Space Institute of Science, 1200 Westlake Ave N Suite 1006, Seattle, WA 98109, USA
}

\date{Version of \today}

\abstract{
Theoretical arguments indicate that close-in terrestial exoplanets may have weak magnetic fields.
As described in the companion article (Paper I), a weak magnetic field results in a high flux of galactic cosmic rays to the top of the planetary atmosphere.}
{We investigate effects that may result from a high flux of galactic cosmic rays
both throughout the atmosphere and at the planetary surface.
}
{Using an air shower approach, we calculate how the atmospheric chemistry and temperature change
under the influence of galactic cosmic rays for Earth-like (N$_2$-O$_2$ dominated) atmospheres. We evaluate the 
production and destruction rate
of atmospheric biosignature molecules. We derive planetary emission and transmission spectra to study the influence of galactic cosmic rays on biosignature detectability. We then calculate the resulting surface UV flux, the surface particle flux, and the associated equivalent biological dose rates.}
{We find that 
up to 20\% of stratospheric ozone is destroyed by cosmic-ray protons. The effect on the planetary spectra, however, is negligible. The reduction of the planetary ozone layer leads to an increase in the weighted surface UV flux by two orders of magnitude under stellar UV flare conditions.
The resulting biological effective dose rate is, however, too low to strongly affect surface life.
We also examine the surface particle flux: For a planet with a terrestrial atmosphere (with a surface pressure of 1033 hPa), a reduction of the magnetic shielding efficiency
can increase the biological radiation dose rate by a factor of two, which is non-critical for biological systems. For a planet with a weaker atmosphere (with a surface pressure of 97.8 hPa), the planetary magnetic field has a much stronger influence on the biological radiation dose, changing it by up to two orders of magnitude.}
{
For a planet with an Earth-like atmospheric pressure, weak or absent magnetospheric shielding against galactic cosmic rays has 
little effect on the planet. It has a modest effect on atmospheric ozone, a weak effect on the atmospheric spectra, and a non-critical effect on biological dose rates. 
For planets with a thin atmosphere, however, magnetospheric shielding controls the surface radiation dose and can prevent 
it from increasing to several hundred times the background level.
}


\titlerunning{Galactic cosmic rays on extrasolar Earth-like planets: II.}
\authorrunning{J.--M. Grie{\ss}meier et al.}

\maketitle

\section{Introduction}

The number of known extrasolar planets is steadily growing, as is the number of known Earth-like 
and ``Super-Earth'' like planets (i.e.~planets with a mass $M\le 10 M\E$) around M-dwarf stars. 
Recent estimations based on the Kepler Input Catalog indicate that the occurrence rate of planets with a radius $0.5R \E \le R \le 4 R\E$ orbiting an M dwarf in less than 50 days is $0.9^{+0.04}_{-0.03}$ planets per star \citep{Dressing13}.
A considerable number of these planets ($\sim$ 15 to 50\%) could be located in the so-called liquid water habitable zone of their host star.

However, the ``classical'' definition of the habitable zone \citep[e.g.,][]{Kasting93,Selsis07} is based solely on the potential of having liquid water on the planetary surface (hence the more precise name of ``liquid water habitable zone'').
Clearly, a number of additional factors can also play an important role for habitability \citep[see e.g.,][]{Lammer09AARV, Lammer10AB}.

One of these additional conditions probably is the presence of a planetary magnetic field.
Magnetic fields on super-Earths around M-dwarf stars are likely to be weak and  short-lived in the best case or even non-existent in the worst case. The relevance of such fields and their potential detectability is discussed elsewhere \citep[][]{Griessmeier14inbook}. 
Here, we look at one habitability-related consequence of a weak planetary magnetic field, namely the enhanced flux of 
galactic cosmic-ray (GCR) particles to the planetary atmosphere, with potential implications ranging from changes in the atmospheric chemistry to an increase of the radiation dose on the planetary surface.

The effects of GCRs on the planetary atmospheric chemistry were calculated by \citet{Grenfell07AB} for N$_2$-O$_2$ atmospheres.
They found that for an unmagnetized planet, GCRs can reduce the total ozone column by almost 20\%, which is not sufficent to strongly influence the biomarker signature. Thus, they concluded that biomarkers are robust against GCRs for the scenarios studied.

GCRs also lead to a flux of secondary particles which can reach the planetary surface. The resulting surface radiation dose was evaluated by \citet{Atri13}. For an Earth-like atmosphere with a surface pressure of 1033 hPa, they find that the absence of magnetospheric shielding can increase the surface biological dose rate by up to a factor of $\sim$ 2. They also indicate that atmospheric shielding dominates over magnetospheric shielding; compared to a thin atmosphere (10 times less dense than on Earth), an atmosphere of 1033 hPa reduces the dose rate by almost 3 orders of magnitude.

The exact severity of these effects, however, depends on the particle energy range considered, and on the intrinsic planetary magnetic field strength. For planets with a strong magnetic field, most galactic cosmic-ray particles are deflected, whereas for weakly magnetized planets, the majority of the particles can reach the planetary atmosphere.
In previous work \citep[][]{Griessmeier05AB,Griessmeier09}, the flux of galactic cosmic rays to the atmosphere of extrasolar planets has been evaluated on the basis of a simple estimate for the planetary magnetic moment. 
However, such quantitative estimates of magnetic fields 
can be over-simplistic.
More complex approaches, however, yield values which are not only model-dependent, but also depend on the precise planetary parameters. For this reason, we have re-evaluated the cosmic-ray flux in the companion article \citep[][hereafter ``Paper I'']{Griessmeier15}, including primary particles over a wider energy range. More importantly, we 
now take a more general approach concerning the planetary magnetic moment: Instead of
applying a model for the planetary magnetic moment, we showed how magnetic protection
varies \textit{as a function of the planetary magnetic dipole moment}, in the 
range $0.0 \,\mathcal{M}\E \le \mathcal{M} \le 10.0 \,\mathcal{M}\E$ for the magnetic moment, 
and in the range of $16 \,\text{MeV} \le E \le \,524 \, \text{GeV}$ for the particle energy. 

The aim of the current study is to use this greatly expanded parameter range and repeat the analysis of earlier studies \citep{Grenfell07AB,Atri13}.
In addition to the new cosmic-ray fluxes from Paper I,
the main differences with respect to the approach of previous work \citep{Grenfell07AB,Atri13} 
are the following:
\begin{itemize}
	\item 	The climate-chemistry atmospheric model has been updated, as explained in Section \ref{sec-model-atmosphere}.
	\item	The calculation of the photochemical response to cosmic rays has been updated, as summarized in Section \ref{sec-model-atmosphere} \citep[see][for details]{TabatabaVakili15}.
	\item	We have added the analysis of planetary transmission and emission spectra (Section \ref{sec-biomarker}).
	\item	We have added the analysis of surface UV flux, and present results for UV-A, UV-B, 
		and biologically weighted UV flux on the planetary surface (Section \ref{sec-uv}).
\end{itemize}	

\begin{figure}[tb] \begin{center}
     \includegraphics[width=0.95\linewidth]
{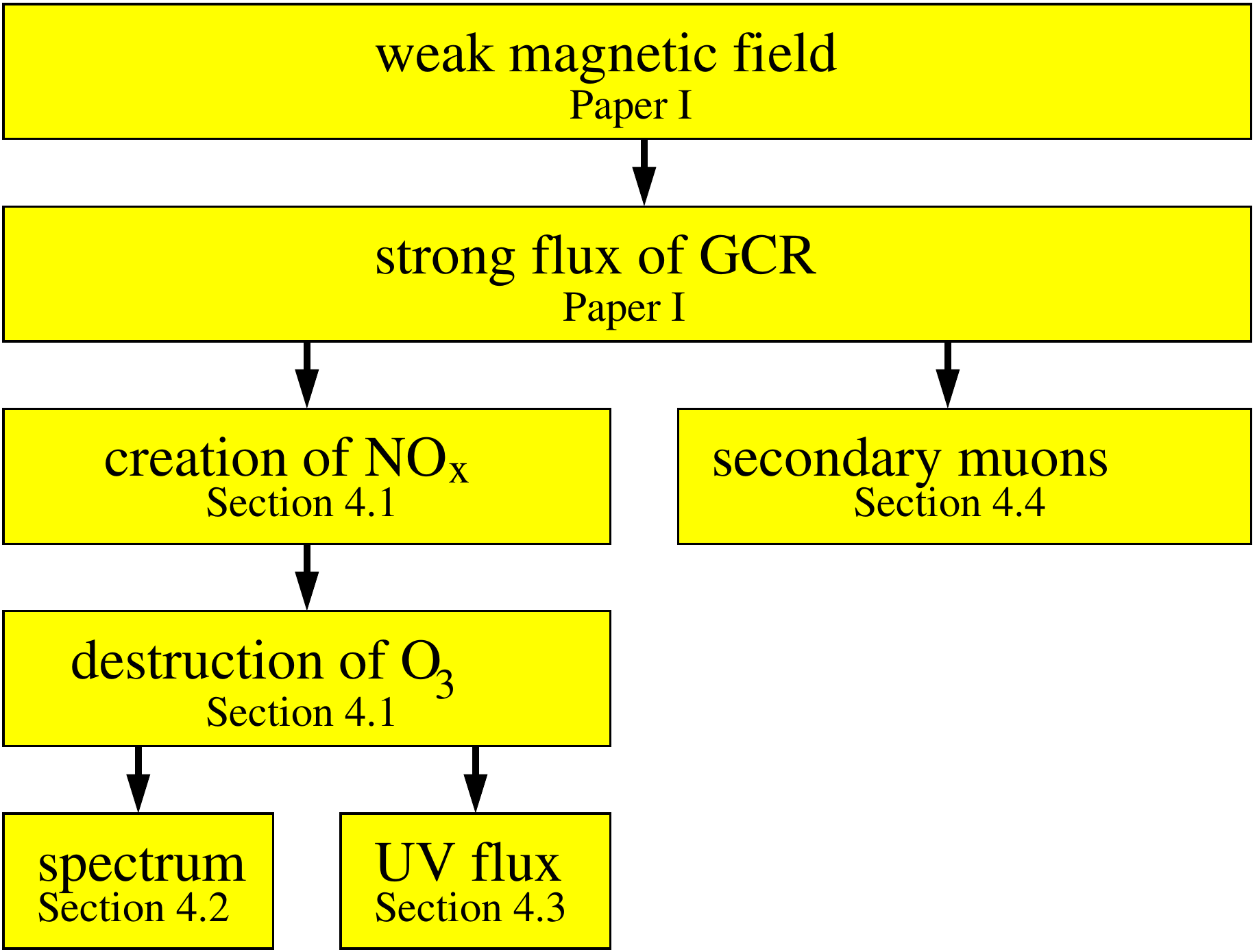}
\caption{Effects of galactic cosmic rays discussed in this work and in the companion article (Paper I).
\label{fig-plan}}
\end{center}
\end{figure}

This paper is organized as follows (see also Figure \ref{fig-plan}):
In Section \ref{sec-planet}, we present the planetary parameters used in our calculations.
Section \ref{sec-model} decribes the models and numerical tools we use: 
The atmospheric chemistry model is discussed in Section \ref{sec-model-atmosphere} and the 
surface particle flux and radiation dose calculation
is presented in Section \ref{sec-model-muon}.
The galactic cosmic-ray fluxes are computed in Paper I.
We discuss the implications of these fluxes in  
Section \ref{sec-implications}, i.e.~the
modification of the atmospheric chemistry (Section \ref{sec-chemistry}), the detectability of biosignature molecules (Section \ref{sec-biomarker}), the UV flux at the planetary surface 
(Section \ref{sec-uv}) and the surface radiation dose rate
(Section \ref{sec-dose}).
Section \ref{sec-conclusions} closes with some concluding remarks.

\section{The planetary situation} \label{sec-planet}

In Paper I,
we calculated a large number of representative cases. This allows us to study systematically the influence of the planetary magnetic field on the flux of GCRs to the planet. 
In this way, we explore the range $0.0 \,\mathcal{M}\E \le \mathcal{M} \le 10.0 \,\mathcal{M}\E$ for the magnetic moment, and the range of $16 \,\text{MeV} \le E \le \,524 \, \text{GeV}$ for the particle energy.

The following parameters are kept fixed:
the stellar mass $M\sstar$ ($M\sstar=0.45 M_{\sun}$), the stellar radius $R\sstar$ ($R\sstar=0.41 R_{\sun}$), 
and the orbital distance $d$ ($d=0.153$ AU). We are thus looking at the case of a planet in orbit around a star equivalent to AD Leonis (an M4.5 dwarf star). 
In addition, we keep constant the planetary mass $M\p$ ($M\p=1.0 M_{\oplus}$) and the planetary radius $R\p$ ($R\p=1.0 R_{\oplus}$).
The only planetary or stellar parameter which we varied in the present study is the planetary magnetic dipole moment $\mathcal{M}$. The minimum value of the magnetic dipole moment in this study is 0, which corresponds to an unmagnetized planet.
The maximum value of the magnetic dipole moment we study is 10 times the present Earth value, which corresponds to an extremely strongly magnetized planet.

Already for Earth-mass rocky exoplanets, a variety of atmospheric compositions can be envisaged. Super-Earths may have a very different atmospheric chemistry from Earth. 
Key processes expected to influence their atmosphere include: the origin of the primary atmosphere, the presence (or absence) of life, 
the presence (or absence) of a water ocean, atmospheric escape and the importance of outgassing \citep[see, e.g.,][]{Hu12,Hu13,Hu14}. 
Having a different atmospheric composition obviously has an effect on cosmic-ray transport and influence the atmospheric profiles of the biomarker molecules we study.
In this work, we assume an Earth-like atmosphere of 1033 hPa surface pressure with N$_2$ and O$_2$ as the major constituents, and with 
biogenic gas emissions as on modern Earth.
The planetary atmospheric parameters are described in more detail in Section \ref{sec-model-atmosphere} and in \cite{Rauer11}.

\section{Models used}
\label{sec-model}

In this section, we describe the models used throughout this article. 
The stellar wind model, magnetic field model, and cosmic-ray propagation model are presented in 
Paper I.
The particle fluxes calculated in Paper I
are used as input into a coupled climate--chemistry atmospheric column model. This model uses an air shower approach to estimate
GCR-induced photochemical effects and the resulting modification of atmospheric chemistry, as described in Section \ref{sec-model-atmosphere}. 
The cosmic-ray fluxes of Paper I are also used as input for the calculation of the 
particle flux to the surface.
We describe our 
surface particle flux and radiation dose model
in Section \ref{sec-model-muon}. 
The results obtained with these models are given in Section \ref{sec-implications}.

\subsection{Climate-chemistry atmospheric column model}
\label{sec-model-atmosphere}

To study the response of the atmosphere to the cosmic-ray flux, we use a Coupled Climate-Chemistry Atmospheric Column Model. The details of this model are decribed elsewhere \citep[][and references therein]{Rauer11,Grenfell12,Grenfell13}.
Since \citet[][]{Grenfell07AB} we include a new offline binning routine for the input stellar spectra and a variable
vertical atmospheric height \citep[see][]{Rauer11}. The code has two main modules, namely, a
radiative-convective climate module (Section \ref{sec-module-climate}) and a chemistry module (Section \ref{sec-module-chemistry}). 
The photochemical response induced
by cosmic rays is modeled according to \citet{Grenfell07AB}, including some model updates described in \citet{TabatabaVakili15} (Section \ref{sec-module-cosmic-rays}).
Finally, the output of the atmospheric model is fed into a theoretical spectral model (Section \ref{sec-module-spectrum}).

The modification of atmospheric chemistry by galactic cosmic rays for our configurations are described in Sections \ref{sec-chemistry}, \ref{sec-biomarker} and \ref{sec-uv}.

\subsubsection{Climate module} 
\label{sec-module-climate}

The \textit{Climate Module} is a global-average, stationary, hydrostatic atmospheric column model ranging from the surface up to altitudes with a pressure of $6.6\cdot 10^{-5}$ bar (for the modern Earth this corresponds to a height of $\sim$70 km). Starting values of composition, pressure, and temperature are based on modern Earth. 
The radiative transfer
is based on the work of \citet[][]{Toon89} for the shortwave region, and on the 
RRTM \citep[Rapid Radiative Transfer Module,][]{Mlawer97} for thermal radiation. 
This uses 16 spectral bands by applying the correlated k-method for major absorbers. Its validity range 
\citep[see][]{Mlawer97} for a given height corresponds to Earth's modern mean temperature $\pm$30K for pressures between $10^{-5}$ and 1.05 bar and for a CO$_2$ abundance from modern up to 100 times modern. The shortwave radiation scheme features 38 spectral intervals for the main absorbers, including Rayleigh scattering for N$_2$, O$_2$, and CO$_2$ with cross-sections based on 
\citet[][]{Vardavas84}. The climate scheme uses a constant, geometrical-mean, solar-zenith angle of 60$^\circ$. 
In the troposphere, lapse rates are derived assuming moist adiabatic convection and using the
Schwarzschild criterion.
Tropospheric humidity comes from Earth observations \citep{Manabe67}. 

For the reference case of the Sun, we employed a high resolution solar spectrum based on \citet[][]{Gueymard04} binned to the wavelength intervals employed in the photochemistry and climate schemes of the column model.

For the standard M dwarf scenario (the ``chromospherically active'' case of Section \ref{sec-uv}), we assume a stellar spectrum identical to that of AD Leonis (an M4.5 dwarf star).
The spectrum is derived from observations of
the IUE satellite and photometry in the visible \citep[][]{Pettersen89}, using observations in the near IR \citep[][]{Leggett96} 
and based on a nextGen stellar model spectrum for wavelengths
beyond 2.4 microns \citep[][]{Hauschildt99ApJ512}.

We also study scenarios with stellar UV flares (``long flare'' and ``short flare'' scenarios). In those cases, the stellar UV spectrum was taken from \citet[][Fig.~3, bold blue line, scaled for distance]{Segura10}. These cases are described in detail in Section \ref{sec-uv}.

Clouds are not included directly, although they are considered in a straightforward manner by adjusting surface albedo to achieve a mean surface temperature of the modern Earth (288 K). 

After convergence, the climate module outputs the temperature, water abundance and pressure.
These variables are interpolated from the climate grid (52 levels) onto the chemistry grid (64 levels)  (both grids extend from the
surface up to about the mid-mesosphere) and are then used as start values for the chemistry module. This, in turn
runs to convergence and then outputs and interpolates the concentrations of key radiative species CH$_4$, H$_2$O, O$_3$ and N$_2$O
to be used as start values for the next cycle of the climate module. This process is repeated back and forth between the
climate and chemistry modules until overall convergence is reached.

\subsubsection{Chemistry module}
\label{sec-module-chemistry}

The \textit{Chemistry Module} has been detailed in 
\citet[][]{Pavlov02}. Our stationary scheme has 55 species for more than 200 reactions with chemical kinetic data taken from the Jet Propulsion Laboratory (JPL) Report \citep[][]{Sander03JPL}. 
Molecules are photolyzed in 108 spectral intervals from 175.4-855nm with an additional nine intervals from 133-173nm and a tenth
interval in the Lyman-alpha. The original chemistry scheme is described in \citet{Kasting84a,Kasting84b}. Further information can also be found at:
http://vpl.astro.washington.edu/sci/AntiModels/models09.html where the original source code is available.
For the absorption cross-sections of key species undergoing photolysis, 
	NO is based on \citet{Cieslik73},
	N$_2$O$_5$ on \citet{Yao82},
	NO$_2$ on \citet{Jones73}, 
	O$_3$ was taken from \citet{Malicet95} and \citet{Moortgat78},
	and NO$_3$ was taken from \citet{Magnotta80}.

We assume a planet with an Earth-like development, i.e.~N$_2$--O$_2$ dominated atmosphere, a modern Earth biomass, etc. The scheme reproduces modern Earth's atmospheric composition with a focus on biosignature molecules (e.g., O$_3$, N$_2$O) and major greenhouse gases such as CH$_4$. The module calculates the converged solution of the standard 1D continuity equations using an implicit Euler scheme. Mixing occurs via Eddy diffusion coefficients (K) based on Earth observations \citep{Massie81}. Constant surface biogenic (e.g., CH$_3$Cl, N$_2$O) and source gas (e.g., CH$_4$, CO) emissions were employed based on the modern Earth 
\citep[see][for more details]{Grenfell11}. 
H$_2$ was removed at the surface as detailed in \citet{Rauer11}.
Also calculated are modern-day tropospheric lightning emissions of nitrogen monoxide (NO), volcanic sulphur emissions of SO$_2$ and H$_2$S, and a constant downward flux of CO and O at the upper boundary, which represents the photolysis products of CO$_2$. Dry and wet deposition is included for long-lived species via deposition velocities (for dry deposition) and Henry's law constants (for wet deposition).

\subsubsection{Cosmic-ray scheme}
\label{sec-module-cosmic-rays}

For the \textit{Cosmic Ray Scheme}, we use an air shower approach based on 
\citet{Grenfell07AB,Grenfell12} and \citet{TabatabaVakili15}. 
The top of atmosphere (TOA) time-average proton fluxes from the magnetospheric cosmic-ray model (Paper I) are input into the chemistry module at the upper boundary.
Secondary particles are generated, which leads to NO$_x$ production.
In the present work 
our scheme was updated to produce 1.25 
odd nitrogen atoms per ion pair produced by cosmic rays, according to \citet{Jackman80}, 
based on calculations of dissociation branching
ratios from relativistic particle impact cross sections \citep{Porter76}.
We introduced a parameterization whereby the GCR-induced N-production was split into two channels, i.e 45\% ground-state N and 55\% excited-state N
\citep[see][and references therein]{Jackman05}. We also introduced an energy-dependence to the total N$_2$ ionization cross section by electron impact \citep{TabatabaVakili15},
replacing the constant electron impact cross section of 1.75 $\cdot$ 10$^{-16}$ cm$^2$ previously used with the energy-dependent cross section of \citet{Itikawa06}.
Additionally, the input parameters for the Gaisser-Hillas formula were extended up to 524 GeV to be consistent with the cosmic-ray calculation. 
Finally, in the Gaisser-Hillas scheme the parametrization was changed. The parameter of the 
\textit{proton attenuation length} (80 g/cm$^2$) was replaced with a \textit{depth of first interaction} of 5 g/cm$^2$ according to 
\citet{AlvarezMuniz02}, which led to a closer match with observations. 
For more details of the above updates and their effects see 
\citet{TabatabaVakili15}. 
Note that the cosmic-ray scheme in the current work includes production of only
nitrogen oxides in the photochemistry (whereas in a newly-developed version as described by
\citet{TabatabaVakili15} the cosmic rays lead to the production of both
nitrogen oxides and hydrogen oxides).

\subsubsection{Theoretical spectral model}
\label{sec-module-spectrum}

To calculate the spectral appearance of our model atmospheres, we use the SQuIRRL code
\citep[Schwarzschild Quadrature InfraRed Radiation Line-by-line,][]{Schreier01}. 
This code was designed to model radiative transfer with a high resolution in the IR region for a spherically symmetric atmosphere 
(taking arbitrary observation geometry, instrumental field of view, and spectral response function into account). 
The scheme assumes local thermodynamic equilibrium; for each layer, a Planck function is used to determine the emission. 
Cloud and haze free conditions without scattering are assumed. 
Absorption coefficients are calculated 
using molecular line parameters from the HITRAN 2008 database \citep{Rothman09},
and emission spectra are calculated 
assuming a pencil beam at a viewing angle of 38$^{\circ}$ as used, for example, in 
\citet{Segura03}.
SQuIRRL has been validated e.g.~by \citet[][]{Melsheimer05}.

\subsection{Surface particle flux and radiation dose calculation}
\label{sec-model-muon}

If particles from a cosmic-ray shower reach the planetary surface, 
biological systems on the planetary surface can be
strongly influenced and even damaged by this secondary radiation. 
In order to assess the expected biological damage, we simulate the air shower and its passage through the atmosphere.
In the case of Earth, muons contribute 75\% of the equivalent dose rate at the surface \citep{OBrien96}, so that our focus lies on these particles, but the contribution of neutrons and electrons are included, too.

Cosmic ray propagation in the atmosphere is a challenging problem, beyond the scope of analytical tools because one has to compute a variety of hadronic and electromagnetic interactions occurring in the atmosphere. Therefore, 
as a complement to the cosmic-ray air shower model which we use for secondary electrons (see Section \ref{sec-module-cosmic-rays}),
we also use a robust Monte Carlo package, CORSIKA v.6990 \citep{Heck98,Heck12software},
which is widely used to simulate air showers for major particle detection experiments.  
The code makes use of a number of packages to model high and low energy hadronic interaction processes and all electromagnetic interactions of charged particles. We take the input cosmic-ray spectrum calculated for different magnetic field cases (the output of Paper I) and model particle propagation with 20 million primary particles for each case. Using such a large ensemble of particles is  necessary to reduce the numerical error as much as possible. Hadronic interactions up to 80 GeV were modeled using the GHEISHA model and above 80 GeV using the SIBYLL 2.1 high-energy hadronic interaction model. None of the ``thinning'' options were used so that no particle information was lost. The final output gives the momentum and types of particles hitting the ground for 20 million primaries. Primaries are incident at the top of the atmosphere from random angles with energies falling randomly according to the energy spectrum. The electromagnetic interactions enhance the atmospheric ionization rate and change the atmospheric chemistry \citep{Atri10JCAP}. For energies of the primary particles above 8 GeV, hadronic interactions produce particles (such as muons and neutrons), some of which reach the ground and contribute to the radiation dose \citep{Atri11,Atri13}. 

The surface radiation calculation thus is similar to the 
calculations of \citet[][]{Atri13}. However, we extend the primary particle energy range and increase the range of planetary magnetic fields. We use an Earth-like atmospheric composition, but compare two different values of atmospheric depth: 1036 g/cm$^2$ (equivalent to a surface pressure of 1033hPa), and 100 g/cm$^2$ (equivalent to a surface pressure of 97.8 hPa). The resulting 
surface radiation equivalent dose rate
is described in Section \ref{sec-dose}.

\section{Implications}
\label{sec-implications}

The interaction of GCR particles with a planet and its atmosphere can lead to a  host of interesting effects, several of which have been suggested to be potentially relevant for habitability. 
The excellent review by \citet[][]{Dartnell11} mentions effects as diverse as:
The {modification of the atmospheric chemistry} (which we discuss below),
the excitation and ionization of atomic and molecular species,
the creation of an ionosphere, 
ions driving atmospheric chemistry and potentially weather and climate dynamics,
the possible influence on atmospheric lightning,
the production of organic molecules within the atmosphere, 
the {destruction of stratospheric ozone} (see below),
the {possible sterilization of the planetary surface} (see below),
and the {degradation of biosignatures} (see below).

In the present work, we focus on the following effects: 
Based on the atmospheric model of Section \ref{sec-model-atmosphere}, we
look at the modification of the atmospheric chemistry (Section \ref{sec-chemistry}), verify the stability of biosignature molecules against destruction by cosmic rays (Section \ref{sec-biomarker}), and analyze the enhanced UV flux resulting from a weakened ozone layer (Section \ref{sec-uv}). Finally, with the 
surface radiation calculation of
Section \ref{sec-model-muon},
we study the biological radiation dose at the planetary surface (Section \ref{sec-dose}).

\subsection{Modification of atmospheric chemistry}
\label{sec-chemistry}

After having traversed the planetary magnetosphere, the galactic cosmic-ray protons reach the planetary atmosphere. 
On the way through the atmosphere, they interact with neutral gas particles, and create secondary electrons via  
impact ionization:
\begin{equation}
	\mathrm{p^+ + X \rightarrow p^+ + X^+ + e^-}. 
	\label{eq-chem-1}
\end{equation}

These free electrons break $\mathrm{N}_2$ molecules, which leads to the formation of NO$_x$: 
\begin{align}
	\mathrm{N_2 + e^-} & \rightarrow \mathrm{2 N + e^-},   \label{eq-chem-2a}\\
	\mathrm{N + O_2} & \rightarrow \mathrm{NO + O}. \label{eq-chem-2b}
\end{align}

Depending on local conditions and the dominating ozone production mechanism, these NO molecules can either destroy ozone (if ozone was created by the Chapman mechanism), or may create ozone (if ozone  is formed via a smog mechanism). 
While both mechanisms are included in our model, ozone production by a Chapman mechanism dominates
in the case of a planet orbiting a chromospherically active M-dwarf star, especially at altitudes above 20 km \citep[][]{Grenfell13}, so that in our case NO mostly destroys stratospheric ozone by catalytic cycles. 
Their results suggest that for planets orbiting chromospherically active M-dwarf stars, 
this process is dominant over most other pathways of ozone destruction, which are also included in our model \citep[or ozone loss via CO oxidation, see][]{Grenfell13}.
The case of catalytic HO$_x$ created by cosmic rays is investigated in more detail elsewhere \citep[][]{TabatabaVakili15}. 
Thus, we have \citep{Crutzen70}:
\begin{align}
	\mathrm{NO+O_3} & \rightarrow \mathrm{NO_2 + O_2} \label{eq-chem-3a}\\
	\mathrm{NO_2 + O} & \rightarrow \mathrm{NO + O_2} \label{eq-chem-3b}\\
	\mathrm{\line(1,0){50}} & \mathrm{\line(1,0){70}} \nonumber \\
	 \text{net: } \mathrm{O+ O_3} & \rightarrow \mathrm{2 O_2}. \label{eq-chem-3net}
\end{align}
The O atom in Eq.~\ref{eq-chem-3b} is created e.g.~by photolysis of O$_2$ 
by photons in the Herzberg region of $\sim$180 nm.

Equation (\ref{eq-chem-3net}) describes the net reaction of the catalytic cycle: Ozone molecules are 
transformed into molecular oxygen. NO$_x$ is regenerated, so that a single molecule 
can contribute to the destruction of a large number of ozone molecules. 

In order to quantify the effects described by equations (\ref{eq-chem-1}) to (\ref{eq-chem-3net}) for extrasolar planets around M-dwarf stars, we use the coupled climate-chemistry model described in Section \ref{sec-model-atmosphere}. In the following, we present the results obtained with this model, probing planets with magnetic moments in the range $0.0 \,\mathcal{M}\E \le \mathcal{M} \le 10.0 \,\mathcal{M}\E$.

\begin{figure*}[tb] \begin{center}
     \includegraphics[width=0.95\linewidth] 		
{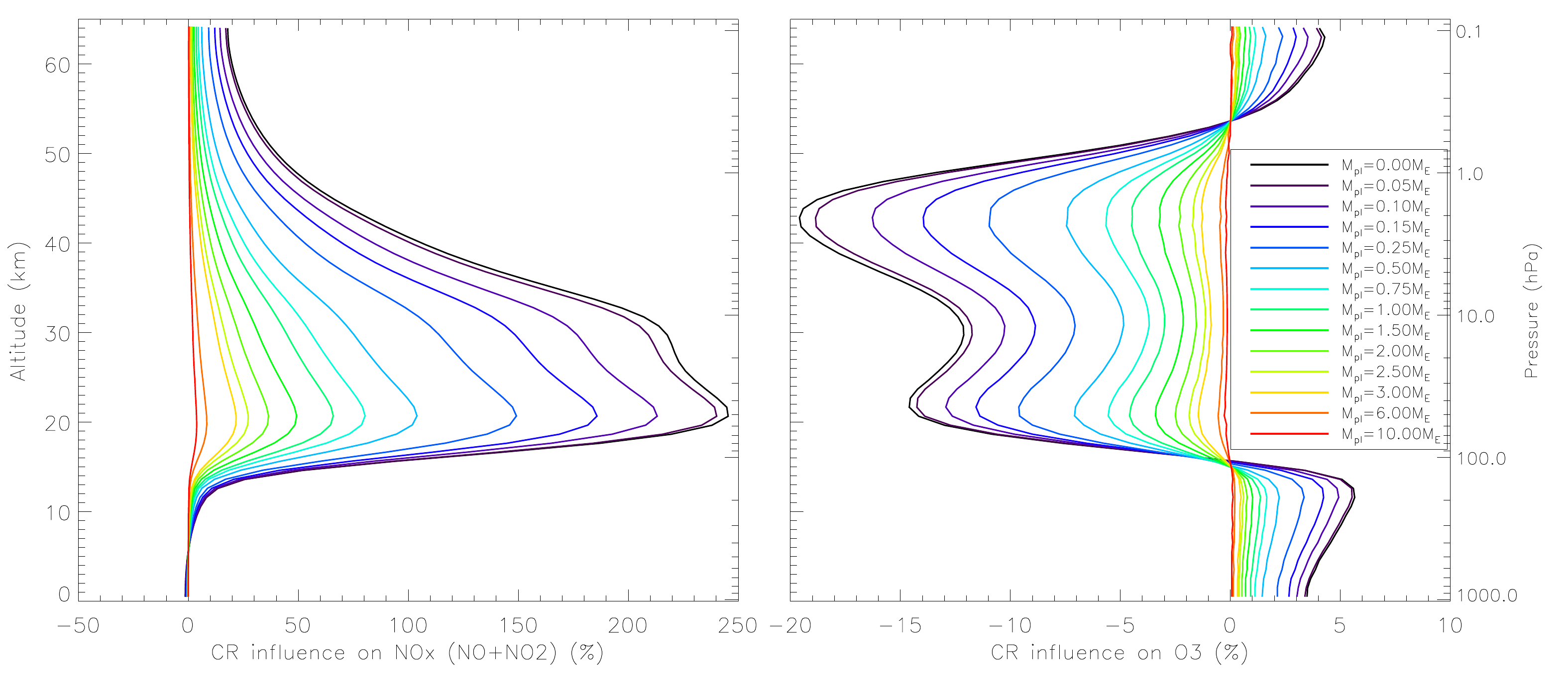}
\caption{Altitude-dependent change of the volume mixing ratio of NO$_x$ (i.e.~NO $+$ NO$_2$) (left) and O$_3$  (right) for exoplanets
with a magnetic moment of $\MM=0.0, 0.05, 0.1, 0.15, 0.25, 0.5, 0.75, 1.0, 1.5, 2.0, 2.5, 3.0, 6.0$ and $10.0\MM_\oplus$, relative to a case without cosmic rays.
\label{fig-atmosphere-profile}}
\end{center}
\end{figure*}

Figure \ref{fig-atmosphere-profile} shows how the influx of galactic cosmic rays into the planetary atmosphere changes the atmospheric compositional profile. 
The left panel of the figure shows the altitude-dependent relative change in the NO$_x$ volume mixing ratio. In the case of a weak magnetic field, the increased influx of galactic cosmic rays enhanced the NO$_x$ by up to a factor 3.5.
This enhancement peaks in the lower stratosphere, where the air shower interaction is strong \citep[e.g.][]{TabatabaVakili15}. 
The right panel of Figure \ref{fig-atmosphere-profile} shows that the increased NO$_x$ 
catalytically destroys up to 20\% of stratospheric O$_3$.
In the troposphere, however, O$_3$ abundances increase by up to 6\% because NO$_x$ stimulates the smog mechanism.

\begin{figure*}[tb] \begin{center}
     \includegraphics[width=0.95\linewidth]
{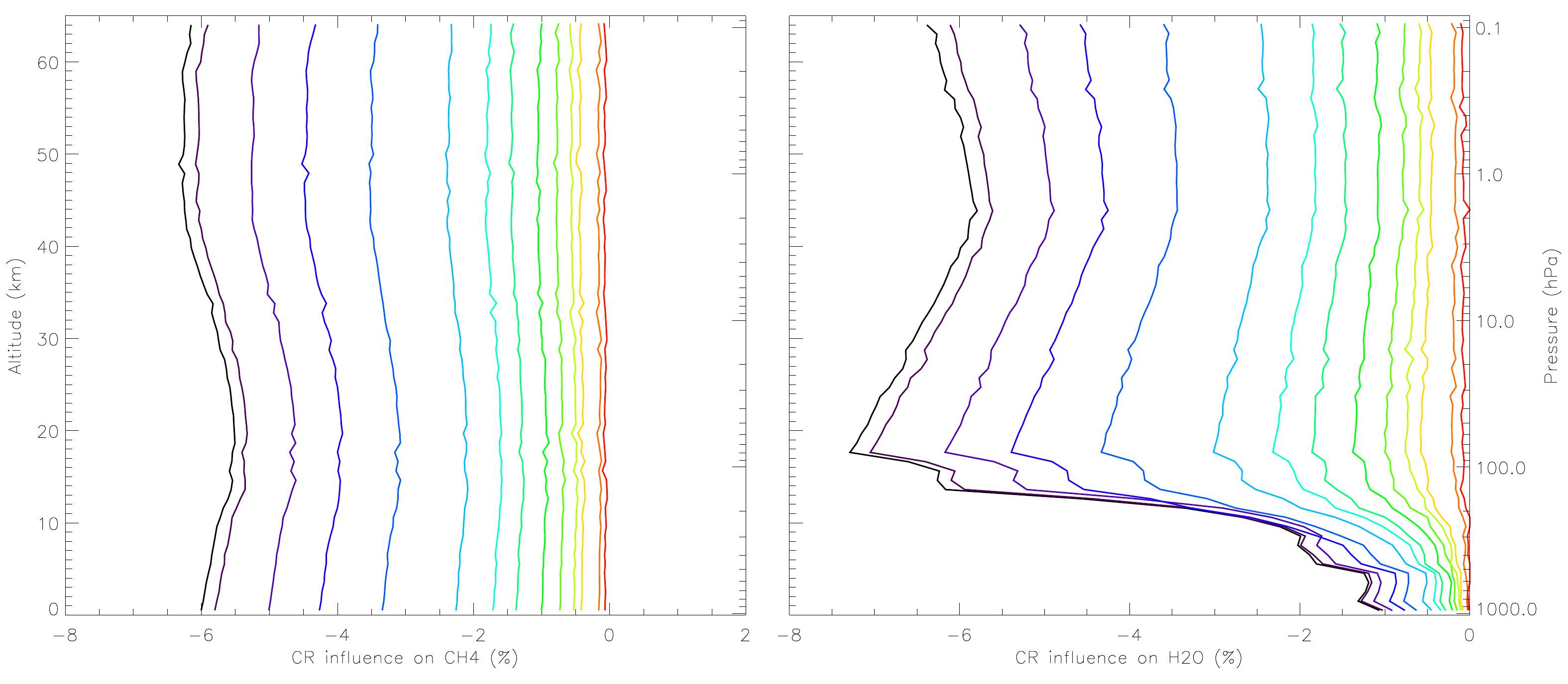}
\caption{
As Fig.~\ref{fig-atmosphere-profile} but for CH$_4$ (left) and H$_2$O (right).
\label{fig-atmosphere-profile-CH4-H2O}}
\end{center}
\end{figure*}

Figure \ref{fig-atmosphere-profile-CH4-H2O} shows the effect of GCRs on CH$_4$ and H$_2$O.
Both chemical species decrease in concentration with decreasing magnetospheric protection (i.e.~increasing GCR flux). 
The left panel shows that the CH$_4$ abundance decreases as shielding decreases.
This is explained by the fact that the increased amount of cosmic-ray particles leads to a higher NO$_x$ abundance, which in turn leads to 
more OH:
\begin{align}
	\mathrm{NO + HO_2} & \rightarrow \mathrm{NO_2 + OH}. \label{eq-rA}
\end{align}

The increased OH abundance leads to a more efficent destruction of CH$_4$:
\begin{align}
	\mathrm{CH_4 + OH} & \rightarrow \mathrm{CH_3 + H_2O}. \label{eq-rB}
\end{align}
Thus, less shielding (i.e. more cosmic rays) leads to less CH$_4$.
A detailed analysis of the runs in the current paper suggests that
OH is the dominant in-situ sink:
Other sinks e.g.~due to
reaction with excited oxygen atoms or with atomic chlorine are weaker and have CH$_4$ removal rates which are lower by at least 
two orders of magnitude. 
When we increase the planetary magnetic moment from 0 to 10 times the Earth's value, the
atmospheric CH$_4$ increased modestly by ~7\%.

The right panel of Figure \ref{fig-atmosphere-profile-CH4-H2O} shows the effect of galactic cosmic rays on the H$_2$O abundance, which decreases
as the magnetic shielding decreases (i.e.~cosmic rays increase).
This is a direct effect of the decrease in CH$_4$, see Eq.~(\ref{eq-rB}). 

\begin{figure*}[tb] \begin{center}
     \includegraphics[width=0.95\linewidth] 
{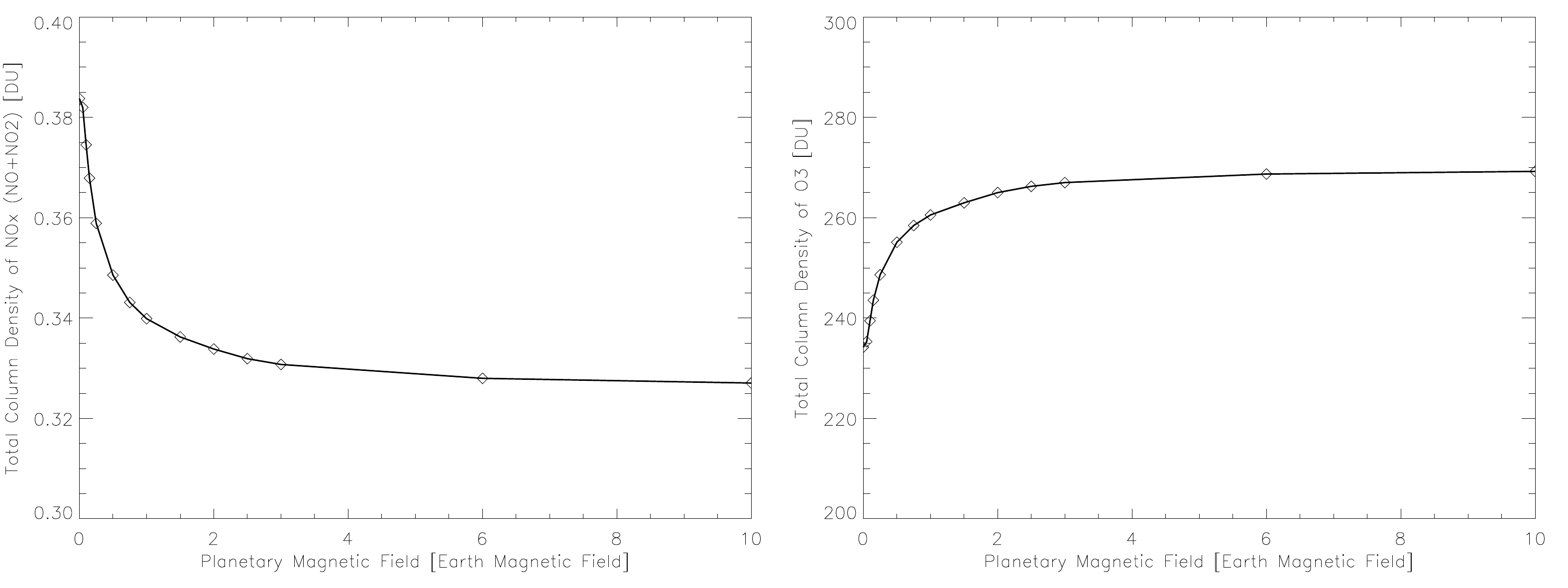}
\caption{Column density of NO$_x$ (i.e.~NO $+$ NO$_2$, left panel) and O$_3$ (right panel) as a function of exoplanetary magnetic moment. 
One Dobson Unit (DU) is equivalent to a column density of 2.69 10$^{16}$ molecules cm$^{-2}$.
\label{fig-atmosphere-column}}
\end{center}
\end{figure*}

Figure \ref{fig-atmosphere-column} shows how the influx of galactic cosmic rays 
changes the total atmospheric column density for NO$_x$ and O$_3$. One can see that under the influence of galactic cosmic rays, the NO$_x$ column varies by up to 15\%, while the O$_3$ column varies by up to 13\%.

If one compares the 0.15 $\MM_\oplus$ case of the present study with the equivalent (run 3)
of our previous work using an earlier model version \citep{Grenfell07AB}, one finds that the results are in good agreement, despite the widened proton energy range and the numerous updates to the routines of the climate-chemistry atmospheric model (see Section \ref{sec-model-atmosphere} for details). 
For example, run 3 of \citet{Grenfell07AB} found 
a 10\% ozone decrease at 30 km altitude, 
and an ozone column value decreasing by 16\%, 
compared to the case without GCR (their run 2). 
In our current model (0.15 $\MM_\oplus$ case), we have 
a 10\% ozone decrease at 30 km altitude 
and an ozone column value decreasing by 10\% compared to a case without cosmic rays. 

Similarly, 
for their run 4 (corresponding to the 0.0 $\MM_\oplus$ case here), \citet{Grenfell07AB} found 
a 12-13\% ozone decrease at 30 km altitude, 
and an ozone column value decreasing by 19\% when compared to 
the case without GCR (their run 2).
In our current model (0.0 $\MM_\oplus$ case), we have 
a 12\% ozone decrease at 30 km altitude, and an ozone column value decreasing by 13\%.
This similarity arises e.g.~from constraints in the maximum $\mathrm{NO}$ production rate by measurement data for the Earth reference case \citep{TabatabaVakili15}.

Absolute values of ozone columns for the AD Leonis scenarios are
lower in the present work than in earlier modeling versions \citep[e.g.][]{Grenfell07AB}. 
Note that in the present work we place the planet at the distance from the star where the net incoming energy equals
one solar constant (instead of a previous approach where the surface albedo was adjusted in
order to reach a surface temperature of 288.0 K).
Other recent model updates are described in \citet{Rauer11}.

In the present work there is a clear smog signal (e.g. Figure \ref{fig-atmosphere-profile}, right panel) leading to an increase of tropospheric O$_3$ when NO$_x$ increases.
This effect was not so evident in earlier studies \citep[e.g.][]{Grenfell07AB}. 
On Earth, smog ozone production has different regimes where it can be sensitive to a) changes in
organic species (e.g., CH$_4$, CO etc.) or b) to changes in NO$_x$. In the present work the absolute CH$_4$ is higher than in 
\citet[][]{Grenfell07AB} (the overall CH$_4$ response is chemically complex, related to changes in OH, see 
\citet[][]{Grenfell12} for some discussion). Higher CH$_4$ is consistent with a saturation of the smog mechanism with respect to organic 
species, hence a more significant role of smog O$_3$ to changes in NO$_x$ -  more work however is required to investigate this 
further.

Similarly to \citet{Grenfell07AB}, we reach the conclusion that atmospheric biosignature molecules are not strongly influenced
by GCRs for Earth-like planets 
in the habitable zone around M-dwarf stars.
The case is, however, different for solar energetic particles, which
can strongly modify the abundance of ozone in the planetary atmosphere
\citep{Segura10,Grenfell12}.
Using the updated model as above, this case is re-evaluated in  
\citet[][]{TabatabaVakili15}.

\subsection{Spectral signature of biosignature molecules}
\label{sec-biomarker}

A particularly useful method to study exoplanets is the analysis of
their atmospheric composition via 
emission or transmission spectroscopy.
In the case of Earth-mass exoplanets or super-Earths, 
a central goal involves the search for
molecules which could indicate the presence of life 
and which cannot be explained
by inorganic chemistry alone, so-called ``biosignature molecules'' (sometimes also called ``biomarkers'').
Several telescopes that 
are currently planned or under construction (e.g., E-ELT, JWST) could possibly detect spectral biosignatures on potentially rocky planets around nearby stars, although this remains very challenging.

Clearly, care has to be taken when selecting and interpreting biosignature molecules.  
Proposed biosignature molecules include oxygen (when produced in large amounts by photosynthesis), ozone
(mainly produced from oxygen) and nitrous oxide (produced almost
exclusively from bacteria).

It is important to understand all effects which can modify the abundances
of these molecules.
For example, \citet{Grenfell07PSS} look at the response of biosignature chemistry (e.g.~ozone) on varying planetary and stellar parameters (orbital distance and stellar type: F, G, and K). \citet{Rauer11} and \citet{Grenfell13} study the effect of stellar spectral type (from M0 to M7, plus the case of the active M-dwarf star AD Leonis, which corresponds to the star used in the present study) and of planetary mass in the Earth to super-Earth range,
whereas \citet{Grenfell14PSS} study the effect of varying stellar UV radiation and surface biomass emissions.

One has to be sure to rule out 
cases where inorganic chemistry can mimic the
presence of life (``false positives'').
Potential abiotic ozone production
on Venus- and Mars-like planets has been discussed by \citet[][and references therein]{Schindler00}. While this is based on photolysis of 
e.g., CO$_2$ and H$_2$O 
 and is thus limited in extent, a sustainable production of abiotic O$_3$
which could build up to a detectable level has been suggested by \citet[][]{DomagalGoldman10} for a planet within the habitable zone of AD Leonis with a specific atmospheric composition. 
Indeed, other studies confirm that abiotic buildup of ozone is possible \citep[e.g.,][]{Hu12,Tian14}; however, detectable levels are unlikely if liquid water is abundant, as e.g.~rainout of oxidized species would keep atmospheric O$_2$ and O$_3$ low \citep[][]{Segura07}, unless the CO$_2$ concentration is high and both H$_2$ and CH$_4$ emissions are low \citep[][]{Hu12}.
False-positive detection of molecules such as CH$_4$ and O$_3$ is discussed by \citet[][]{vonParis11}. 
\citet{Seager13} present a biosignature gas classification.
Since abiotic processes cannot be ruled out for individual molecules (e.g. for O$_3$), searches for biosignature molecules should 
search for multiple biosignature species simultaneously.
It has been suggested that the simultaneous presence of O$_2$ and CH$_4$ can be used as an indication for life \citep[][and references therein]{Sagan93}. Similarly, \citet{Selsis02} suggest a so-called ``triple signature'', where the combined detection of O$_3$, CO$_2$ and H$_2$O would indicate biological activity. \citet{DomagalGoldman10} suggest to simultaneously search for the signature of O$_2$, CH$_4$, and C$_2$H$_6$. 
Of course, care has to be taken to avoid combinations of biosignature molecules which can be generated abiotically together \citep[see e.g.][]{Tian14}.
The detectability of biosignature molecules is discussed, e.g.~by 
\citet[][]{vonParis11} and \citet{Hedelt13}.
In particular, the simulation of 
the instrumental response to simulated spectra
for currently planned or proposed exoplanet characterization missions 
has shown that the amount of information the retrieval process can provide on the atmospheric composition may not be sufficient
\citep[][]{vonParis13}.

Similar to ``false positives'', which can lead to erroneous interpretation of observational data, one also has to deal with the problem of ``false negatives''
for life-bearing planets. The absence of ozone does not necessarily mean that life is absent. Oxygen or ozone may be quickly consumed by chemical reactions, preventing it from reaching detectable levels \citep[][]{Schindler00,Selsis02}.
Also, non-detection can result from masking by a wide CO$_2$ absorption \citep[][]{Selsis02,vonParis11}.
Here, we look into an abiotic process (namely GCRs) which can destroy the signature of biosignature molecules.
Similarly to potential false-positives, these effects have to be taken into account in order to correctly interpret observational data.

As has been shown in Section \ref{sec-chemistry} (Figure \ref{fig-atmosphere-column}), the ozone column
can be modified by up to 13\% 
by the action of GCRs in the case of weak magnetic fields.
We find similar values for other biosignature molecules.
In the following, we explore the question: Could this modify the observed spectrum of a planet, either in emission or in transmission?

\begin{figure*}
\begin{center}
	\subfigure[Planetary emission contrast spectrum.]
	{\includegraphics[angle=0,width=0.45\linewidth]
{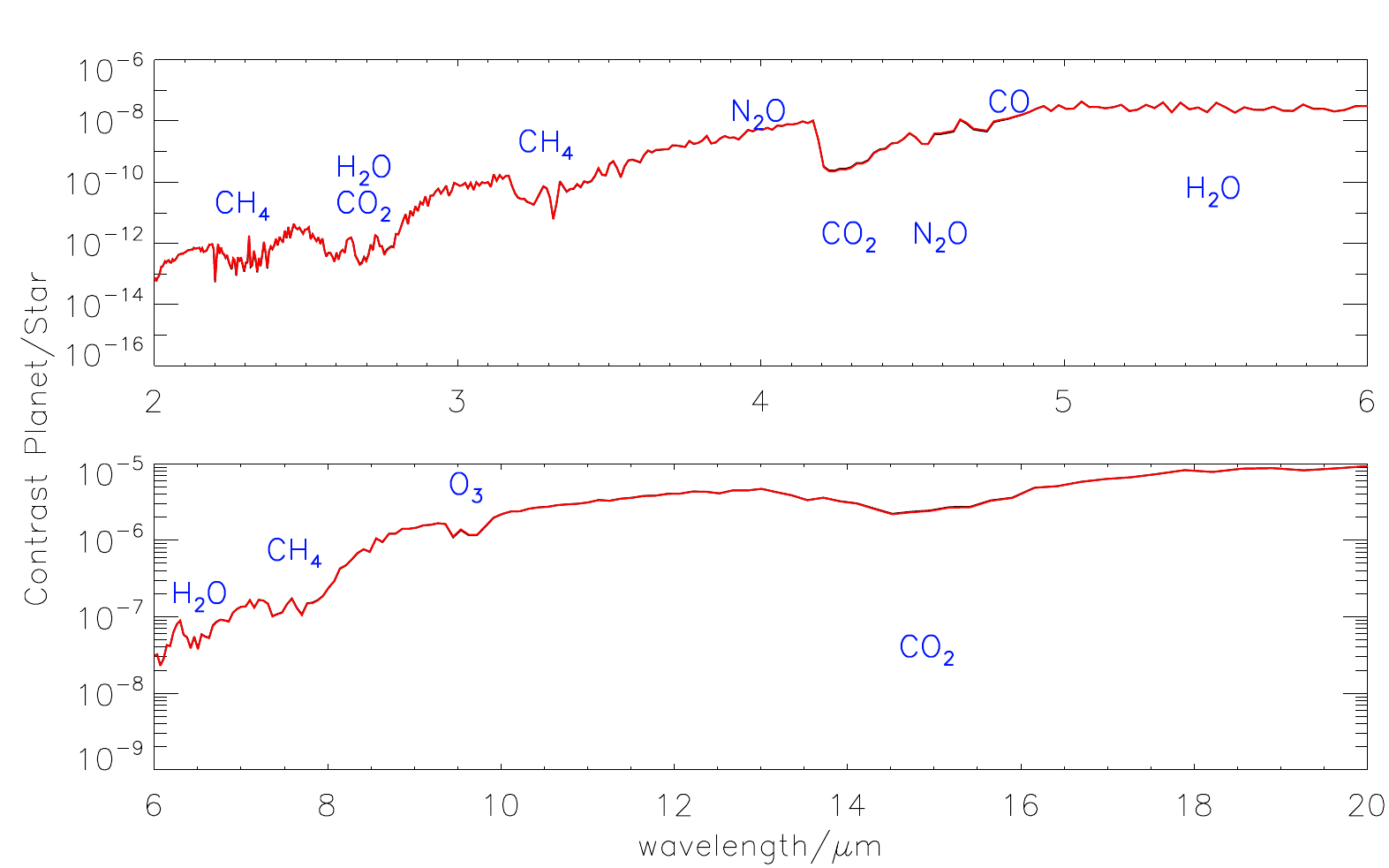}
	\label{fig-atmosphere-spectrum-emission}}
	\subfigure[Relative transmission spectrum.]
	{\includegraphics[angle=0,width=0.45\linewidth]
{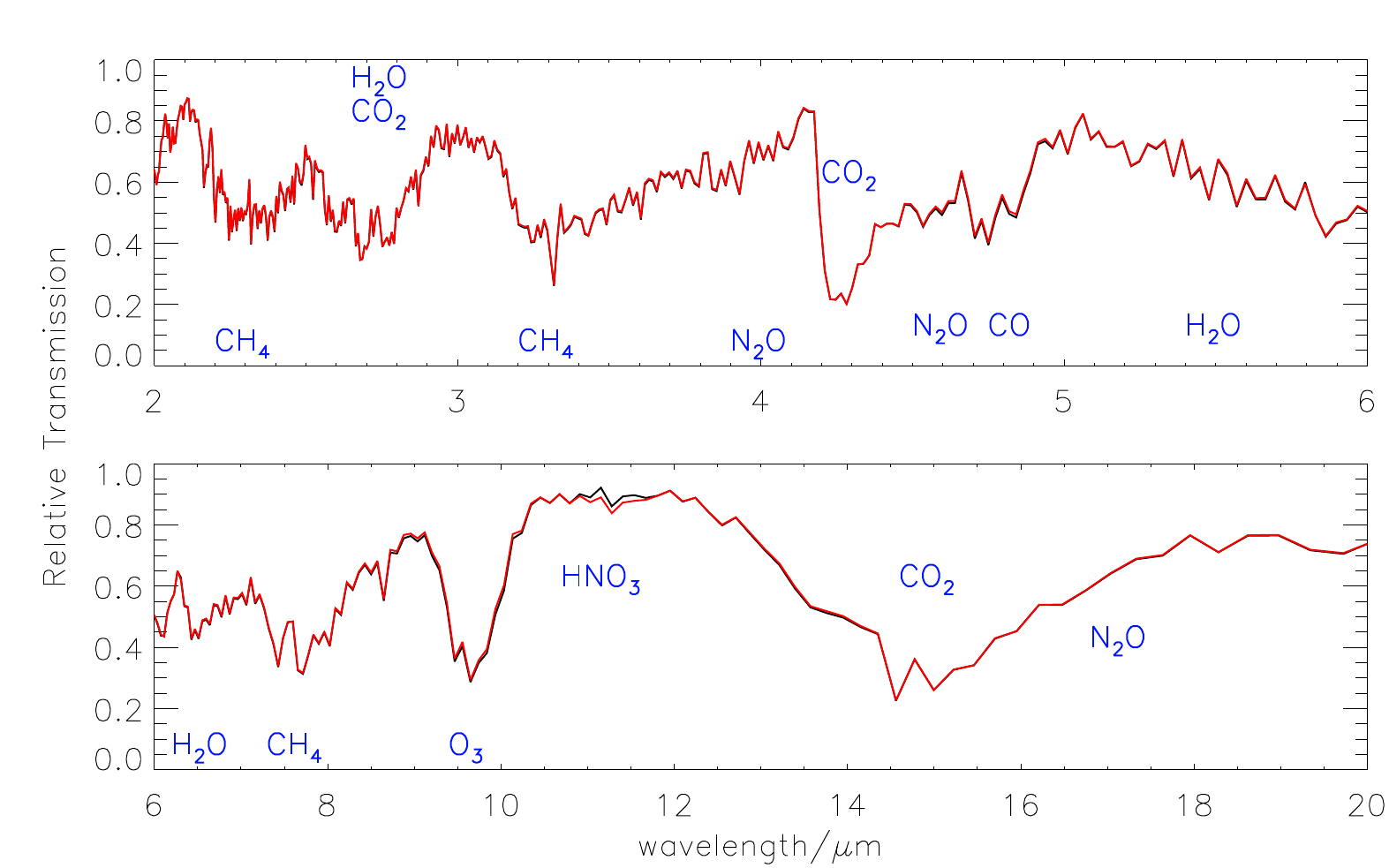}
	\label{fig-atmosphere-spectrum-transmission}}
\caption{Emission contrast spectrum (left) and relative transmission spectrum (right)
	with and without GCRs for an Earth-like planet orbiting in the habitable zone of the M-dwarf star AD Leonis. 
	Black line (mostly hidden under the red line): no GCRs. Red line: maximum GCRs (i.e.~no planetary magnetospheric protection). 
The spectra were calculated for R=1000 (total number of spectral bins).
\label{fig-atmosphere-spectrum}}
\end{center}
\end{figure*}

Figure \ref{fig-atmosphere-spectrum}
explores the influence of GCRs on the molecular signature in the planetary spectrum for wavelengths between $2 \le \lambda \le 20 \mu$m 
using the SQuIRRL code (cf.~Section \ref{sec-module-spectrum}).
Figure \ref{fig-atmosphere-spectrum-emission} shows the planetary emission spectrum, whereas 
Figure \ref{fig-atmosphere-spectrum-transmission} shows the relative transmission coefficient 
\citep[see][for details on the spectral methods]{Rauer11}.
In both figures, the black line corresponds to the reference spectrum, i.e.~the case of a planet orbiting in the habitable zone of an M dwarf star with no GCRs (zero cosmic-ray case), whereas the red line corresponds to the M dwarf scenario with GCR and $\MM=0.0\MM_\oplus$ (i.e.~maximum galactic cosmic-ray case). Note that the red line mostly lies over the black line.
Both figures indicate that spectral observations would show no detectable difference between the cases with and without GCRs. The same is true for observations at higher spectral resolution 
(e.g.~for a total number of spectral bins R=10\,000, not shown).

We thus reach the conclusion that the influence of
GCRs on atmospheric biosignature molecules
for Earth-like planets  
in the habitable zone of M-dwarf stars
is too weak to be detectable in the planetary spectra.
The case of solar energetic particles
is discussed in  
\citet[][]{TabatabaVakili15}.

\subsection{Surface UV flux}
\label{sec-uv}

Besides changes in the planetary spectrum, a direct consequence
of the loss of stratospheric ozone is an increase in the surface UV radiation, especially in the UV-B range. 
We would like to know whether 
this change is important enough to potentially have an impact on 
life (since most surface life relies on a protective O$_3$-layer).

The atmosphere-penetrating UV radiation is frequently divided into three different ranges: 
\begin{itemize}
\item
UV-A, 
(3150 to 4000\AA $\,$ according to the 
definition of the ``Commission Internationale de l'Eclairage'', CIE), 
which has the smallest biological significance, 
\item
UV-B (2800 to 3150\AA $\,$ according to the CIE definition), which is 
biologically damaging,
\item
 and UV-C (1754 to 2800\AA $\,$ in this study), 
which is strongly damaging, but
of which very little reaches the ground for an Earth-like atmosphere. 
\end{itemize}
The biological radiation damage created by UV-A radiation is 5 orders of magnitude weaker than the damage caused by UV-C radiation \citep{Horneck95,Cockell99,Cuntz10IAU}. Even so, UV-A still produces significant mutagenic and carcinogenic effects 
\citep[e.g.,][]{deGruijl00,Scalo07}.
Due to very efficient atmospheric shielding, the surface flux of UV-C is usually many orders of magnitude smaller than that of UV-A or UV-B, even during stellar flares \citep{Segura10}.
For this reason, most work \citep[e.g.][]{Grenfell12} concentrates on UV-B radiation (280-315 nm).
In this work, however, we proceed slightly differently. Similarly to \citet{Segura10}, we study the full UV spectrum from $1754-3150$ \AA, and present integrated results for the two relevant bands UV-A and UV-B. The surface UV-C fluxes turn out to be negligible due to atmospheric shielding.

One should note that the ways in which UV radiation can be harmful for living cells are complex and varied \citep[e.g.,][and references therein]{deGruijl00,Scalo07}. 
In addition, some species have found ways to protect themselves against
harmful radiation, either through repair mechanisms \citep[][]{Scalo07}, protective layers, or through 
strategies which allow to avoid strong radiation altogether \citep[e.g.~Section 7 of][]{Heath99,Scalo07}.
The influence of a highly fluctuating environment on life is discussed by \citet[][and references therein]{Scalo07}.

The UV flux of different M-dwarf stars can differ considerably, and is highly variable in time \citep{France13}.
\citet{Grenfell14PSS} investigate the effect of different stellar UV fluxes on atmospheric biosignatures.
When analyzing the flux of the UV radiation of an M-dwarf star to the planetary surface, we have to distinguish between a number of different cases: The quiescent stellar UV flux of a non-active star, the UV flux of a chromospherically active star (such as AD Leo or GJ 643C), and the UV flux during a stellar UV flare. For the latter case, we differentiate between ``long'' and ``short'' flares, depending on the flare duration relative to the atmospheric response timescale 
(see below). 
Thus, the relevant cases are:
\begin{itemize}
	\item[E)]	We compare to the case of Earth in the habitable 
			zone of the Sun (i.e.~at a distance of 1 AU). 
	\item[Q)]	The \textit{quiescent} emission of (model) non-active stars is characterized by 
		UV fluxes
		many orders of magnitude smaller than for the Sun over the whole UV range 
		\citep{Segura05}. Recent studies indicate 
		that this case may be less representative than previously thought \citep{France13}. 
		No noticeable effect is expected and thus this case is not 
		further discussed in this work.
	\item[CA)]	
		\textit{Chromospherically active} stars have additional flux in the range 100-300 nm, 
		generated by chromospheric and coronal activity. 
		Although their absolute UV flux is inferior to that of the Sun 
		\citep[e.g.][Fig. 1]{Buccino07}, their normalized UV flux (i.e.~at a distance 
		where the surface temperature or total flux equals that on present Earth) 
		however may exceed the solar value 
		\citep[e.g.~by a factor up to 10 in UV-C,][]{Segura05}. 
		Recent studies indicate that AD Leo may be more active
		than previously thought \citep{France13}. 
		For a planet in the habitable zone of a chromospherically active M-dwarf star, we use the top
		of atmosphere (TOA) UV spectrum of AD Leonis, as described in Section 
		\ref{sec-model-atmosphere}.
	\item[LF)]	
		During a stellar flare, the UV flux typically increases by one 
		order of 
		magnitude \citep{France13}, and by at least two orders of magnitude 
		in more extreme
		cases \citep[][who looked at different stages of the 1985 flare of AD Leo]{Scalo07,Segura10}, with 
		a timescale of 10$^2$-10$^3$ seconds. For a planet in the habitable zone of a 
		flaring M-dwarf star, we use the TOA UV spectrum of 
		\citet[][their Fig.~3, bold blue line, scaled for distance]{Segura10}, i.e.~the maximum flux at
		the flare peak. 
		The TOA flux is approximately solar for UV-A and 2.5 times solar for UV-B 
		(see Table \ref{tab-uv-star}). 
		For UV-C, the TOA flux is $\sim 10$ times solar.
		The timescale of the flare has to be compared to the timescale over which the atmosphere responds.
		As we are using a stationary model, we can only probe the extreme cases of very long 
		(quasi-continuous) stellar UV flares, where the time between flares is so short that the atmosphere is 
		constantly under flaring conditions, and of very short (isolated) flares, where the atmosphere does not 
		have the time to react to the flare.
		In the case of a \textit{long flare}, the flare timescale is longer than the typical 
		reaction time of the planetary atmosphere and the atmosphere adjusts to the modified conditions.  
		In this case, we calculate the surface UV flux from the modified TOA flux using the model of Section 
		\ref{sec-model-atmosphere}. 
		This case 
		also applies when the planet is subject to a quasi-continuous succession of UV flares, 
		which might well be the case for planets around active M-dwarf stars 
		\citep{Khodachenko07AB,Grenfell12}.   
	\item[SF)]
		If, on the other hand, the timescale of the UV flare (e.g.~its duration) is short 
		compared to the atmospheric reaction time, the atmosphere has not yet adjusted to the 
 		increased UV flux. In this \textit{short flare} 
		case, the atmosphere, and thus its transmission ratio  
		$R(\lambda)=I_\text{surface}(\lambda) / I_\text{TOA}(\lambda)$ 
		are identical to the pre-flare conditions, i.e.~the case CA described above.
		Hereby, $I_\text{surface}(\lambda)$ and $I_\text{TOA}(\lambda)$ 
		denote the flux at the planetary surface and at the top of the atmosphere, respectively.
		The top-of-atmosphere UV-flux, however, is identical to the flaring case, LF.
		We thus use the transfer function $R(\lambda)$
		obtained in the case CA, and multiply it with the TOA UV flux of the case LF
		\citep[][their Fig.~3, bold blue line  scaled for distance]{Segura10} 
		to obtain the surface UV flux. 
		On Earth and in our exoplanetary calculations, most atmospheric ozone is located below 50 km where reaction 	
		timescales are long \citep[e.g.][]{Allen84}, so that this scenario is appropriate for 			
		an isolated flare (i.e.~the flare timescales are much shorter than the atmospheric reaction timescales).
	\item[SCR)]	
		Stellar UV flares are expected to be frequently accompanied by \textit{stellar cosmic-ray} (SCR) particles, which 
		are not included in the above cases LF and SF.
		The influence of such SCRs (as opposed to the GCRs discussed in the present work) are only  
		briefly discussed below. A more detailed analysis is presented separately 
		\citep[][]{TabatabaVakili15}. 

\end{itemize}
In addition to the UV flux emitted by the planetary host star, another source can contribute to atmospheric and surface UV. \citet{Smith04} have suggested that stellar X-rays may be reprocessed in the atmosphere, generating an additional contribution of UV photons. They found that up to 10\% of the X-ray energy may be redistributed into the UV range by aurora-like emission in the absence of UV-blocking agents 
(i.e.~when the UV transport is defined by Rayleigh scattering alone).
In the case of the Earth, UV redistribution may transfer a fraction of 
$2\cdot10^{-3}$ of the incident energy to the planetary surface in the 200-320 nm range. 
\citet{Segura10} estimated the X-ray energy for a strong flare on the M-dwarf star AD Leo flare to be 9 W/m$^2$, so the energy redistributed as UV radiation at the
planetary surface should be $<0.018$ W/m$^2$. This is negligible compared to the UV flux of the flare itself (cf. Table \ref{tab-uv-star}).

For the above cases E), CA) and LF), we are interested in the transmission of UV radiation through the atmosphere. 
For each case, we investigate the full range from $\mathcal{M}=0$ (i.e.~no magnetospheric shielding, where the atmospheric ozone is most strongly depleted by galactic cosmic rays) up to $10\,\mathcal{M}_{\oplus}$ (i.e.~strong magnetospheric shielding), plus the case without GCRs (which, for our purposes, corresponds to a planet with an infinite magnetic moment, and thus a planet with maximum stratospheric ozone shield).

\begin{figure}
\begin{center}
	{\includegraphics[angle=0,width=0.9\linewidth]
{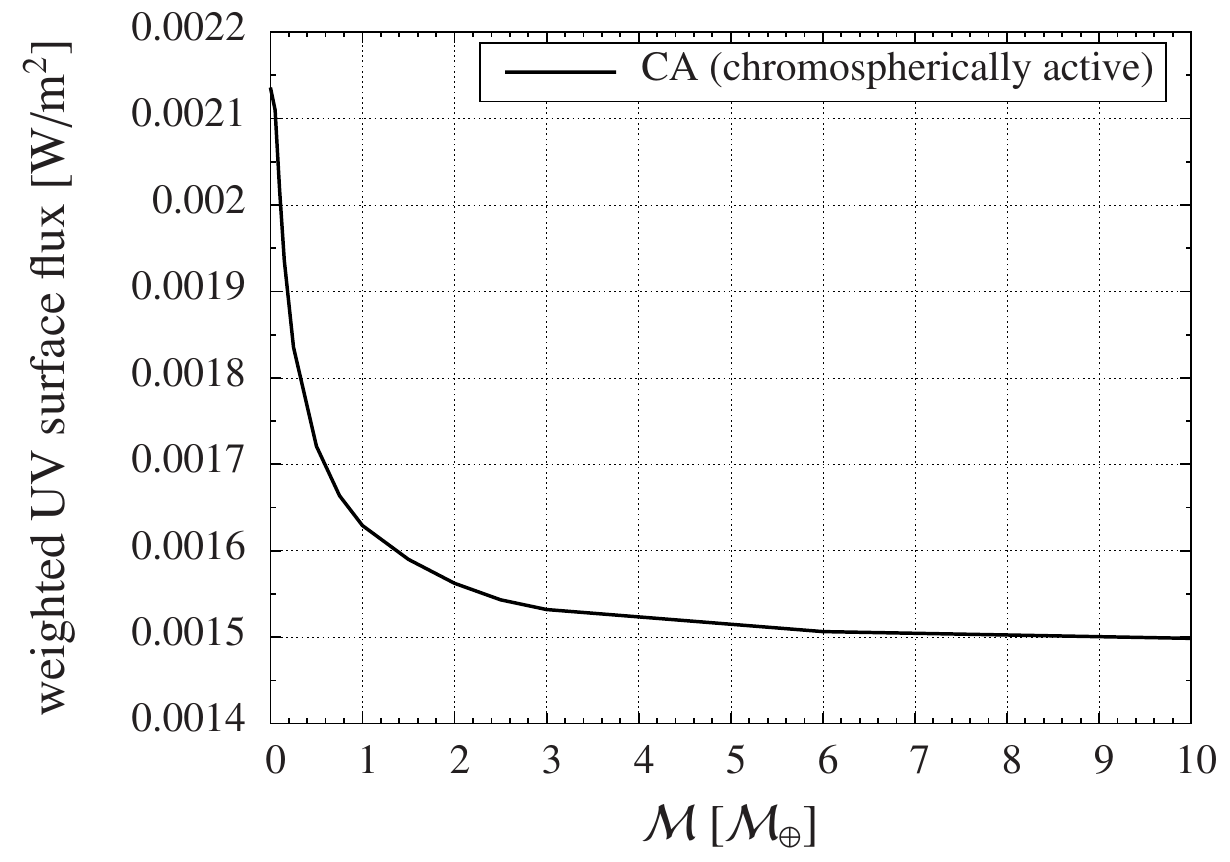}}
\caption{Biologically weighted surface UV flux $W$ as a function of magnetospheric shielding for a chromospherically active star (case CA).
\label{fig-surface-uv-weighted-ca}}
\end{center}
\end{figure}

\begin{figure}
\begin{center}
	{\includegraphics[angle=0,width=0.9\linewidth]
{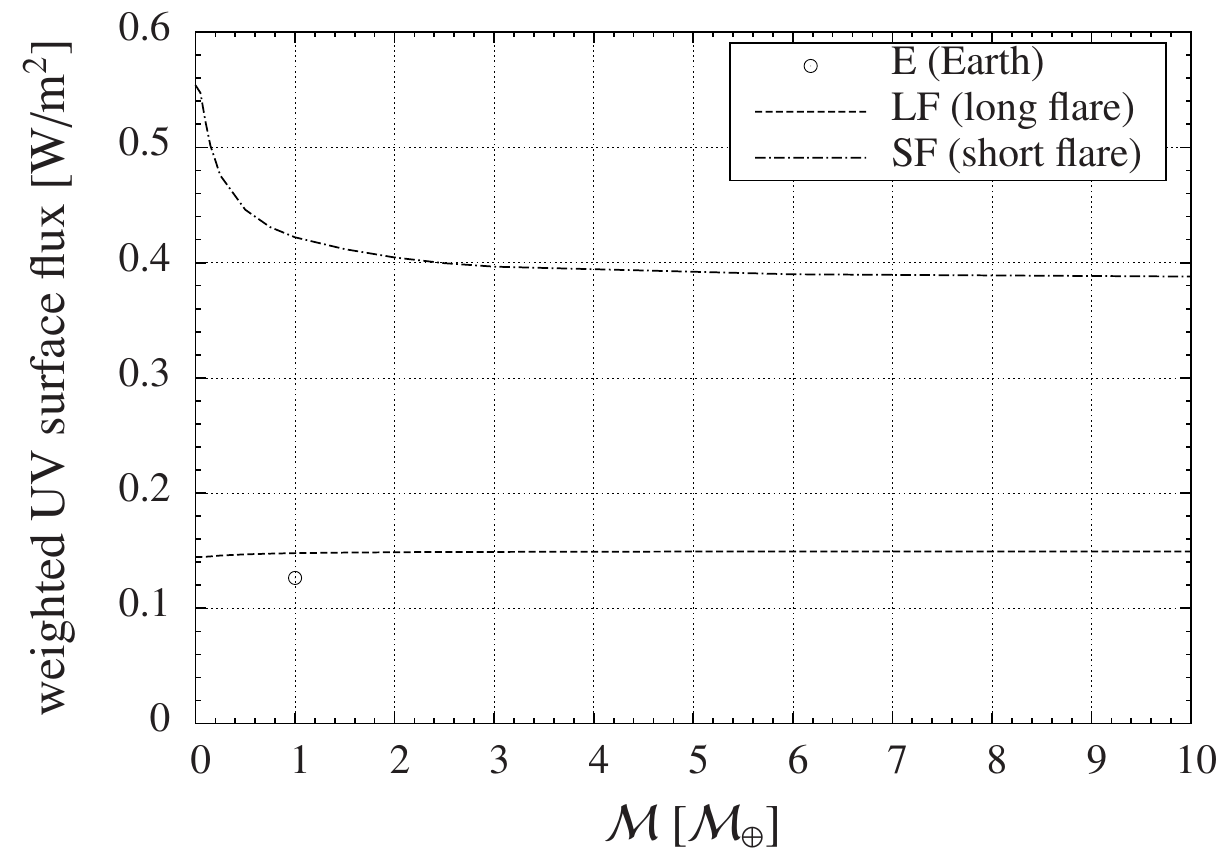}}
\caption{Biologically weighted surface UV flux $W$ as a function of magnetospheric shielding. 
Dashed line: Long flare (case LF).
Dash-dotted line: Short flare (case SF).
Circle: Earth (case E).
\label{fig-surface-uv-weighted-lf-sf}}
\end{center}
\end{figure}

We proceed as follows:
\begin{itemize}
	\item	
		We take the wavelength-resolved stellar UV flux 
		and calculate the flux incident at the top of atmosphere (TOA) of the M-dwarf star planet
		corresponding to the planetary orbital distance.
		From this we calculate the wavelength-integrated TOA UV fluxes  	
		$I^{\text{UV-A}}_\text{TOA}$ and $I^{\text{UV-B}}_\text{TOA}$ 
		(Columns 3 and 6 of Table \ref{tab-uv-star}). 
	\item	
		With this UV flux, the numerical model described in Section \ref{sec-model-atmosphere}, 
		and using the magnetic-field dependent
		TOA fluxes of GCR particles from Paper I, 
		we calculate the wavelength-resolved flux of UV at the planetary surface. 
		From this we calculate the wavelength-integrated surface UV fluxes  
		$I^{\text{UV-A}}_\text{surface}$ and $I^{\text{UV-B}}_\text{surface}$
		(Columns 4 and 7 of Table \ref{tab-uv-star}). 
		The effects that are considered here are absorption (by O$_3$ and other species) and Rayleigh 
		scattering by atmospheric 
		molecules. 
	\item	
		We calculate the ratio of UV penetrating through the planetary atmosphere, averaged over 
		the corresponding UV band,
		e.g.~$R^{\text{UV-A}}=I^{\text{UV-A}}_\text{surface} / I^{\text{UV-A}}_\text{TOA}$, 
		and similarly for $R^{\text{UV-B}}$ 
		(Columns 5 and 8 of Table \ref{tab-uv-star}). 
		$R$ thus characterizes the average UV shielding by the atmosphere in a particular 
		wavelength band.
		Note that the value of $R$ depends on both the TOA GCR flux and the TOA 
		UV flux (the atmosphere behaves as a non-linear system).    
	\item
		We multiply the wavelength-resolved UV surface spectra with the DNA action spectrum of 
		\citet[][Figure 1]{Cuntz10IAU}, which is based on previously published data \citep[][]
		{Horneck95,Cockell99}, to calculate the effective biological UV flux $W$ at 
		the planetary surface (Column 9 of Table \ref{tab-uv-star}, and 
		Figures \ref{fig-surface-uv-weighted-ca} and \ref{fig-surface-uv-weighted-lf-sf}).
		In this, the DNA action spectrum is normalized to 1 at a wavelength of 300 nm 
		(i.e.~at 300 nm, a flux of 1 W/m$^2$ contributes 1 W/m$^2$ to $W$).
		We also compare the relative contribution of UV-A, UV-B and UV-C to the effective 
		biological UV flux $W$.
\end{itemize}	

\begin{table*}[!h]
\begin{center} 
\small{
   \begin{tabular}{|l|c||c|c|c||c|c|c||c|}\hline
        Column 1 & Column 2 & 3 & 4 & 5 & 6 & 7 & 8 & 9 \\[4pt] \hline \hline
        case  & $\MM$ 
		& $I^{\text{UV-A}}_\text{TOA}$ & $I^{\text{UV-A}}_\text{surface}$ & $R^{\text{UV-A}}$
		& $I^{\text{UV-B}}_\text{TOA}$ & $I^{\text{UV-B}}_\text{surface}$ & $R^{\text{UV-B}}$
		& $W$ 
		\\[4pt] 
	        & [$\MM_\oplus$] 
		& [W/m$^2$] & [W/m$^2$] &
		& [W/m$^2$] & [W/m$^2$] &
		& [W/m$^2$] 
		\\[4pt] \hline \hline
	E & 1.0 
		& $127$ & 90.41 & 0.71
		& $18.29$ & $2.264$ & $0.12$
		& 0.126
		\\[4pt] \hline  	
	E & no GCR
		& $127$& 90.41 & 0.71
		& $18.29$ & $2.256$ & $0.12$
		& 0.125
		\\[4pt] \hline	
	\hline
	CA & 0.0
		& 2.01 & 1.479 & 0.74
		& $0.202$ & $0.0225$ & 0.11
		& 0.0021
		\\[4pt] \hline
	CA & 0.1 
		& 2.01 & 1.479 & 0.74
		& $0.202$ & $0.0221$ & 0.11
		& 0.0020
		\\[4pt] \hline
	CA & 1.0  
		& 2.01 & 1.477 & 0.74
		& $0.201$ & $0.0204$ & 0.10
		& 0.0016
		\\[4pt] \hline
	CA & 10.0  
		& 2.01 & 1.477 & 0.74
		& $0.200$ & $0.0198$ & 0.10
		& 0.0015
		\\[4pt] \hline 	
	CA & no GCR      
		& 2.01 & 1.477 & 0.74
		& $0.200$ & $0.0197$ & 0.10
		& 0.0015
		\\[4pt] \hline
	\hline
	LF & 0.0  
		& $112$ & 75.75 & 0.68
		& 47 & 3.053 & 0.065
		& 0.145
		\\[4pt] \hline
	LF & 0.1  
		& $112$ & 75.75 & 0.68
		& 47 & 3.054 & 0.065
		& 0.145
		\\[4pt] \hline
	LF & 1.0  
		& $112$ & 75.79 & 0.68
		& 47 & 3.084 & 0.066
		& 0.148
		\\[4pt] \hline
	LF & 10.0  
		& $112$ & 75.80 & 0.68
		& 47 & 3.099 & 0.066
		& 0.149
		\\[4pt] \hline	
	LF & no GCR     
		& $112$ & 75.81 & 0.68
		& 47 & 3.100 & 0.066
		& 0.150
		\\[4pt] \hline
	\hline	      
	SF & 0.0  
		& $112$ & 76.50 & 0.68
		& 47 & 5.10 & 0.11
		& 0.55
		\\[4pt] \hline
	SF & 0.1  
		& $112$ & 76.47 & 0.68
		& 47 & 5.00 & 0.11
		& 0.52
		\\[4pt] \hline
	SF & 1.0  
		& $112$ & 76.37 & 0.68
		& 47 & 4.65 & 0.10
		& 0.42
		\\[4pt] \hline
	SF & 10.0  
		& $112$ & 76.33 & 0.68
		& 47 & 4.52 & 0.10
		& 0.39
		\\[4pt] \hline	
	SF & no GCR     
		& $112$ & 76.33 & 0.68
		& 47 & 4.51 & 0.10
		& 0.39
		\\[4pt] \hline
	\hline	      
   \end{tabular}
   \caption[]
   {UV-A and UV-B flux at top of atmosphere ($I^{}_\text{TOA}$) 
   and the surface ($I^{}_\text{surface}$) for different cases. 
   Also shown: wavelength-range averaged flux ratio 	
   $R^{}=I^{}_\text{surface}/I^{}_\text{TOA}$ (i.e.~atmospheric 
   transmission coefficient), and biologically weighted surface UV flux $W$ (in weighted W/m$^2$).
   Cases: E = Earth, CA = chromospherically active star, LF = long flare, SF = short flare
   (see text for details).\\
   }
\label{tab-uv-star}
}
\end{center} 
\end{table*}

Our main results are described in the following (see also Table \ref{tab-uv-star}).
\paragraph{Case E (Earth)}: 
In the case of the Earth, GCRs leave both the UV transmission coefficients and the UV surface fluxes virtually unchanged. As a consequence, the biologically weighted UV surface flux $W$ is barely affected by the presence of GCRs (Table \ref{tab-uv-star}, Column 9). 

The surface UV-B results, i.e.~the surface flux (Table \ref{tab-uv-star}, Column 7) and the transmission coefficient (Table \ref{tab-uv-star}, Column 8) show a good accordance with \citet[][Table 2, line 1]{Grenfell12}, where $R^\text{UV-B}=0.13$ for the case with GCRs, and with \citet{Grenfell13}, where $R^\text{UV-B}=0.16$ for the case without GCRs. These values 
are also compatible with Earth observations \citep[][Table 2, line 4]{Grenfell12}.

For the biologically weighted flux, we find $W=0.126$ W/m$^2$, which is dominated by the contribution of UV-B (94\%), with a small contribution from UV-A (6\%). The influence of UV-C on $W$ is negligible. 
\citet{Cockell99} obtain a lower value for $W$, which is possibly due to their stronger ozone layer.

\paragraph{Case CA (chromospherically active star)}: 
In the case of a planet in the habitable zone of a chromospherically active M-dwarf star, the transmission ratio for UV-A 
is $R^{\text{UV-A}}=0.74$ (Table \ref{tab-uv-star}, Column 5), independent of magnetic shielding, and similar to the case of the Earth.

As shown in Table \ref{tab-uv-star}, 
the cosmic-ray induced weakening of the ozone layer described in Section \ref{sec-chemistry} has little influence on the atmospheric UV-B transmission ratio $R^{\text{UV-B}}$ (Column 8), which increases from 0.10 to 0.11 with decreasing magnetic shielding (i.e.~increasing GCR effect). Between strong and zero magnetic shielding, the UV-B surface flux increases by 14 \% (for a decrease of the ozone Column by 13\%, see Section \ref{sec-chemistry}). 
This is consistent with the near-linear relationship between atmospheric ozone column and surface UV-B flux observed on Earth \citep[e.g.][]{Kerr93}.
Due to the low intensity of UV-B emitted by the M-dwarf star, the resulting UV-B surface flux is very small (two order of magnitude less than on present-day Earth, see Column 7). 

The surface UV-C flux (not shown) is negligibly small, even when the high biological response factor to UV-C is taken into account.

Figure \ref{fig-surface-uv-weighted-ca} shows how the biologically weighted UV surface flux $W$ changes as a function of magnetic shielding; between minimum and maximum magnetic shielding, $W$ changes by $\sim 40$\%. 
It is dominated by the contribution of UV-B ($\ge95$\%), with a small contribution from UV-A (3-5\%). The influence of UV-C on $W$ is negligible. 
A detailed analysis shows that for our parameters the UV-B flux at 300 nm has the strongest contribution to $W$. $W$ is considerably lower than on Earth (case E; see Column 9 of Table \ref{tab-uv-star}).
Our results are in good agreement with \citet{Grenfell13}, who find $R^\text{UV-B}=0.11$ for AD Leo (case without cosmic rays).

\paragraph{Case LF (long stellar UV flare)}: 
As shown in Table \ref{tab-uv-star}, the results are different in the case of an exoplanet either exposed to a long UV flare or to a quasi-continuous succession of UV flares (\textit{case LF}).
The UV-A transmission ratio $R^{\text{UV-A}}$ is similar to that of the case CA, but the higher TOA flux leads to an increased surface UV-A intensity.

For UV-B, the transmission ratio is reduced compared to the case CA. 
However, this is compensated by the increased TOA flux, so that the surface flux is higher than for the chromospherically active case (about two orders of magnitude) and exceeds the level of case E.

Again, the surface UV-C flux (not shown) is negligibly small, even when the high biological response factor to UV-C is taken into account.

Compared to the case CA, the biologically weighted UV surface flux $W$ (which is mostly determined by shortwave UV-B) is increased by up to two orders of magnitude in the case LF. It is comparable to the value for Earth (case E, circle in Figure \ref{fig-surface-uv-weighted-lf-sf}. See also Column 9 of Table \ref{tab-uv-star}).
$W$ is dominated by the contribution of UV-B ($\ge92$\%), with a small contribution from UV-A (7-8\%). The influence of UV-C on $W$ is negligible. 

\begin{figure*}[tb] \begin{center}
     \includegraphics[width=0.95\linewidth] 
{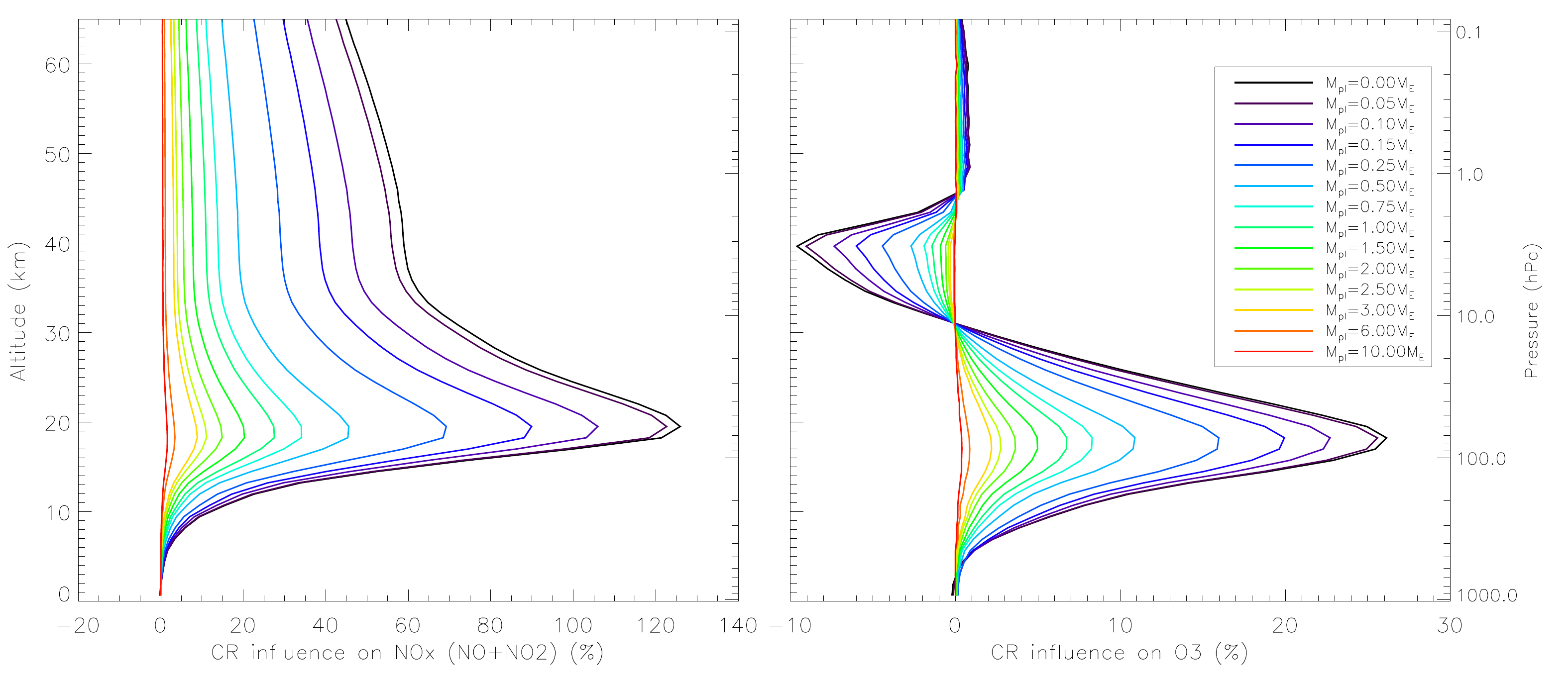}
\caption{
As Figure \ref{fig-atmosphere-profile} but for the case of a long UV flare (LF).
\label{fig-atmosphere-column-LF}}
\end{center}
\end{figure*}

Another effect of long flares can be seen in Table \ref{tab-uv-star} and 
Figure \ref{fig-surface-uv-weighted-lf-sf}: When the magnetic moment increases, i.e.~when the planet is better shielded against GCRs, the surface UV flux shows a slight \textit{increase}! 
This is a consequence of the enhanced stellar UV flux, which leads to extra ozone production 
in the altitude region 5-30 km, with a peak at 18 km (Figure \ref{fig-atmosphere-column-LF}).
This suggests an increasing smog mechanism at low altitudes, whereas above 30 km O$_3$ is still destroyed by catalytic NO$_x$. 
Figure \ref{fig-atmosphere-column-LF} shows that with increasing GCR flux (decreasing magnetic shielding), the smog mechanism in the lower atmosphere dominates over the catalytic destruction in the upper atmosphere, so that the column integrated ozone content increases, and the surface UV flux decreases (Figure \ref{fig-surface-uv-weighted-lf-sf}).

\paragraph{Case SF (short stellar UV flare)}: 
For a short stellar flare (with a timescale of 10$^2$-10$^3$ seconds), we do not have the inversed response 
(i.e.~increase of surface UV with increasing magnetic moment) of case LF, but instead (by construction) a behavior identical to the case CA, with higher absolute flux values (cf. Table \ref{tab-uv-star}). Also, by construction, the averaged UV transmission rates are similar to the case CA and the atmospheric profile is identical to that of the case CA (Figure \ref{fig-atmosphere-profile}). 

As a result of the higher input flux, the surface flux of UV-A and UV-B is at least 50 times higher than in the case CA, see Table \ref{tab-uv-star}.

In the case SF, the biologically weighted UV surface flux $W$ is higher than in the case CA by a factor 200-300.
Again, $W$ is dominated by the contribution of UV-B ($\ge97$\%), with a small contribution from UV-A (2-3\%). The influence of UV-C on $W$ is negligible. 

The modulation by the magnetic field is comparable to the case CA  (compare Figures \ref{fig-surface-uv-weighted-ca} and \ref{fig-surface-uv-weighted-lf-sf}): Between $\mathcal{M}=0$ and $1.0\mathcal{M}_{\oplus}$, $W$ decreases by 30\%. For higher magnetic fields, $W$ continues to decrease, by up to 10\% 
(Figure \ref{fig-surface-uv-weighted-lf-sf}). As indicated in Table \ref{tab-uv-star} (Column 9) and Figure \ref{fig-surface-uv-weighted-lf-sf}, $W$ is a factor of 3-4 higher than on Earth (case E) for the duration of the short flare.

\paragraph{Case SCR (stellar cosmic rays)}: 
UV flares are frequently accompanied by stellar cosmic-ray particles \citep{Segura10}, which would amplify the ozone destruction. In that case, the strong removal of stratospheric ozone \citep{Grenfell12} can reduce the UV shielding to $\sim 50\%$, and lead to considerably  higher UV surface fluxes, surpassing those on the terrestrial surface by an order of magnitude.
Using the current model, this case is re-evaluated in a separate article 
\citep[][]{TabatabaVakili15}. 

\paragraph{Discussion}: 
The comparison shows that a short flare is potentially more harmful than a long flare, in which the atmosphere has time to adjust to the high UV flux and absorption is increased at mid-altitudes.
The effect of GCRs on planetary UV radiation is weaker than the modification caused by a change in the stellar spectrum (e.g.~case E to CA). Looking at the different wavelength ranges, one notices that:
\begin{itemize}
	\item	GCRs leave the UV-A transmission coefficient and flux virtually unchanged. 
	\item	For UV-B, GCRs modify the transmission rate and surface flux by less than 20\%, and the relative change is proportional to the change in ozone column, as expected. 
	\item	The surface flux of UV-C remains negligibly small in all cases.
	\item	GCRs may change the biologically weighted UV surface flux $W$ by up to 40\%.
		In all cases, $W$ is dominated by the contribution of UV-B ($\ge 90$\%), with a minor 	
		contribution from UV-A (2-8 \%). UV-C does not contribute significantly to $W$.
	\item	The GCR-induced variation of $W$ (40\%) is much less than the difference due to a change in stellar emission during a flare. For example, during a short stellar flare (case SF), one finds values 3-4 times higher than on Earth (case E), or 200-300 times the quiescent level (case CA).
	\item   
	 	The GCR-induced UV radiation is $W\le 0.55$ W/m$^2$ (Table \ref{tab-uv-star}, Column 9). This value  
		has to be compared to the tolerance of biological systems.
		In particular, \textit{Deinoccocus radiodurans} 
		is able to withstand high levels of UV radiation. \citet{Gascon95} estimate the D$_{90}$ dose (i.e.~ the dose for inactivation 
		of 90\% of the bacterial population) of \textit{Deinoccocus 	
		radiodurans} to be $\sim 553$ J/m$^{2}$. 
		For their measurements, they used a UV lamp with a flux density of $1.7$ W/m$^{2}$ at 256 nm; 
		the D$_{90}$ time their case was thus of the order of 5 minutes.  
		The flux density of their lamp corresponds to an equivalent DNA effective irradiance of $43$ W/m$^{2}$ \citep{Cockell99}, 
		i.e.~almost two order of magnitude stronger than the maximum
		UV flux caused by GCRs, cf.~Table \ref{tab-uv-star}. The D$_{90}$ time for \textit{Deinoccocus radiodurans} 
		on the surface of a magnetically 
		unshielded exoplanet exposed 
		to GCRs plus stellar UV flares would thus be of the order of 7 hours.
		Also, bacteria are usually not fully exposed. 
		\citet{Cockell99} estimates that life on Earth may have arisen during times 
		when the biologically weighted UV flux was 
		$>96$ W/m$^{2}$, which is higher by a factor 170 than the maximum value for GCR-induced UV ($W\le 0.55$ W/m$^2$, 
		cf.~Table \ref{tab-uv-star}).
		We thus conclude that GCR-induced UV radiation can be considered as non-critical.
\end{itemize}

We conclude that GCR-induced UV radiation is non-critical, 
but we note that the UV environment on M-dwarf star planets is much more variable than on Earth.

\subsection{Surface biological dose rate}
\label{sec-dose}

In this section, we analyze which fraction of cosmic-ray particles can reach the planetary surface, and evaluate the associated biological dose rate.

Like on Earth, the surface of an exoplanet can be shielded against galactic cosmic rays by two barriers. The first barrier is the planetary magnetosphere, which deflects particles provided their energy is low enough (Paper I).
However, this does not mean that all 
particles that penetrate through the magnetosphere
reach the surface. The atmosphere acts as a second barrier, and prevents low-energy particles and their products from reaching the surface \citep{OBrien96}.  
At Earth, the minimum energy a proton must have to initiate a nuclear interaction sequence detectable at the surface is approximately 450 MeV \citep{Shea00}.
Higher energy protons generate an atmospheric nuclear cascade or cosmic-ray shower, with high energy secondary particles such as neutrons, electrons, pions and muons reaching the planetary surface. 
The low-energy components of the cascade are absorbed in the atmosphere, leading to an altitude with maximum particle flux, the Pfotzer maximum. In the case of the Earth, the Pfotzer maximum is located at an altitude of 15-26 km, depending on latitude and solar activity level \citep{Bazilevskaya08}. Below the Pfotzer maximum, the particle flux decreases toward the surface. Depending on the altitude of the Pfotzer maximum, the 
surface radiation dose can either be lower or higher than at the top of the atmosphere. For a planet with an Earth-like atmosphere, the absorption effect dominates, and the atmosphere has to be regarded as a second barrier which partially protects the surface against the 
cosmic-ray flux.

If parts of the cosmic-ray shower reach the planetary surface, 
one could expect that biological systems there can be
strongly influenced and even damaged by this secondary radiation. 
This expectation is backed up by experimental evidence, which shows that during Ground Level Enhancements (extreme events where large numbers of secondary cosmic rays reach the Earth's surface) DNA lesions on the cellular level increase considerably
\citep[][and references therein]{Belisheva05,Griessmeier05AB,Belisheva06,Dartnell11,Belisheva12}.
In the case of Earth, muons contribute 75\% of the equivalent dose rate at the surface \citep{OBrien96}.

In Paper I, we have shown that weakly magnetized super-Earths orbiting M-dwarf stars can be exposed to much higher cosmic-ray fluxes at the top of the atmosphere when compared to the case of the Earth.
The question naturally arises: How does this high flux at the top of the atmosphere translate into a radiation dose at the planetary surface?

The details of this interaction and the resulting radiation dose on the planetary 
surface depend on the planetary atmospheric pressure and composition.
Thus, the best way to address this issue is to simulate numerically
the interactions by following the particles from the top of the atmosphere down to 
the planetary surface.
For this, we use the 
surface particle flux and radiation dose model
as described in Section \ref{sec-model-muon}. First results have been presented by \citet[][]{Atri13}; for the current work, the range of planetary magnetic moments has been extended.

\begin{figure}[tb] \begin{center}
     \includegraphics[width=0.95\linewidth] 
{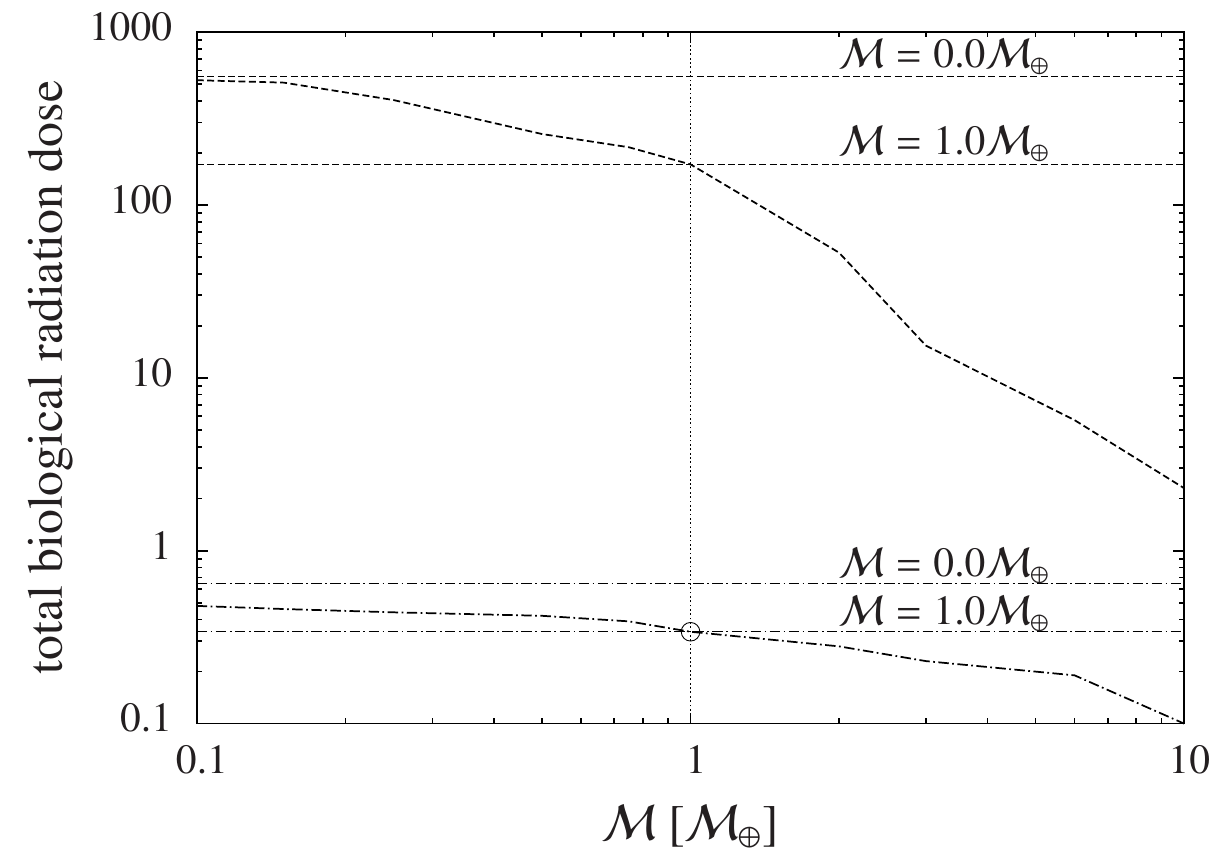}    

\caption{Total biological dose rate (i.e.~the sum of radiation dose rates by  muons, electrons, and neutrons, in mSv/yr) as a function of magnetospheric shielding.
Bold dash-dotted line: planet with an atmospheric depth of 1036 g/cm$^2$ (i.e.~an Earth-like atmosphere with a surface pressure of 1033 hPa). 
Bold dashed line: planet with an atmospheric depth of 100 g/cm$^2$ (i.e.~ a surface pressure of 97.8 hPa). 
Vertical line (shown as a guide for the eye): $\MM=1.0 \MM_\oplus$.
Horizontal lines (shown for comparison): Total biological dose rates for the case $\MM=0$ (upper dashed and upper dash-dotted horizontal line) and $\MM=1.0 \MM_\oplus$ (lower dashed and lower dash-dotted line).
Circle: Earth.
\label{fig-surface-dose}}
\end{center}
\end{figure}

\begin{table}[!h]
\begin{center} 
   \begin{tabular}{|c||c|c|c|}\hline
        $\MM$ [$\MM_\oplus$] 
		& $B^{\text{1036}}$ [mSv/yr] & $B^{\text{100}}$ [mSv/yr]
		\\[4pt] \hline \hline
	0.0
		& 0.65 & 553
		\\[4pt] \hline
	0.1 
		& 0.48 & 527
		\\[4pt] \hline
	0.15 
		& 0.46 & 510
		\\[4pt] \hline
	0.25
		& 0.44 & 405
		\\[4pt] \hline
	0.5
		& 0.42 & 257
		\\[4pt] \hline
	0.75
		& 0.39 & 216
		\\[4pt] \hline
	1.0  
		& 0.34 & 172
		\\[4pt] \hline
	2.0  
		& 0.28 & 53
		\\[4pt] \hline
	3.0  
		& 0.23 & 15
		\\[4pt] \hline
	6.0
		& 0.19 & 5.7
		\\[4pt] \hline
	10.0  
		& 0.1 & 2.3 
		\\[4pt] \hline 	
	\hline	      
   \end{tabular}
   \caption[]
   {Total biological dose rate $B$ (i.e.~the sum of radiation dose rates by  muons, electrons, and neutrons, in mSv/yr) in the case of
    modified magnetospheric shielding for a planet with an atmospheric depth of 1036 g/cm$^2$ ($B^{\text{1036}}$) and for 
    a planet with an atmospheric depth of 100 g/cm$^2$ ($B^{\text{100}}$).\\
   \label{tab-cr-dose}
   }
\end{center} 
\end{table}

Figure \ref{fig-surface-dose} and Table \ref{tab-cr-dose} show the total biological radiation dose rate as a function of the planetary magnetic field, measured in mSv/yr.
In Figure \ref{fig-surface-dose}, the dash-dotted line corresponds to a planet with an atmospheric depth of 1036 g/cm$^2$ ($B^{\text{1036}}$, i.e.~the biological radiation dose rate for an Earth-like atmosphere with a surface pressure of 1033 hPa), while the dashed line corresponds to a planet with an atmospheric depth of 100 g/cm$^2$ ($B^{\text{100}}$, the biological radiation dose rate for a planet with a surface pressure of 97.8 hPa). The vertical line denotes Earth's magnetic moment ($\MM=1.0 \MM_\oplus$), and the circle indicates Earth-like conditions. The horizontal lines are shown to guide the eye. They indicate the total biological dose rates for the cases $\MM=0$ 
(upper dashed and upper dash-dotted horizontal line) and $\MM=1.0 \MM_\oplus$ (lower dashed and lower dash-dotted horizontal line).

In the case of a planet with an Earth-like atmosphere with a surface pressure of 1033 hPa (dash-dotted line),
magnetospheric shielding 
reduces the surface biological dose rate by a factor of approximately 2 between $\mathcal{M}=0$ and $1\,\mathcal{M}_{\oplus}$. Obviously, for stronger magnetic fields ($\mathcal{M}>1\,\mathcal{M}_{\oplus}$), the biological dose rate further decreases (by another factor of 3  for $\MM=10.0 \MM_\oplus$).
As the magnetic field decreases, the filter efficiency of the magnetosphere decreases, and the number of cosmic-ray protons reaching the top of the planetary atmosphere increases (Paper I). However, the atmosphere remains as a second filter, and removes most of the biologically relevant particles,
so that 
the total biological radiation dose rate of Figure \ref{fig-surface-dose} (i.e.~the sum of the radiation dose rates by muons, electrons, and neutrons) increases only slowly with decreasing magnetic moment.

For a 
planet with a weaker atmosphere having a surface pressure of 97.8 hPa (dashed line), magnetospheric shielding is more important, and reduces the surface biological dose rate by a factor of approximately 3 between $\mathcal{M}=0$ and $1\,\mathcal{M}_{\oplus}$, and another factor of 70 between $\mathcal{M}=1$ and $10\,\mathcal{M}_{\oplus}$.

As was already noted by \citet[][]{Atri13}, atmospheric shielding dominates over magnetospheric shielding. In 
Figure \ref{fig-surface-dose} and Table \ref{tab-cr-dose},
this is indeed obvious: At $\mathcal{M}=0$, the Earth-like atmosphere 
(bold dash-dotted curve in Figure \ref{fig-surface-dose}) reduces the surface biological dose rate by almost three orders of magnitude when compared to the weak atmosphere case (bold dashed curve). For an Earth-like magnetic moment ($\mathcal{M}=1\,\mathcal{M}_{\oplus}$), the difference is still more than two orders of magnitude.
For strongly magnetized planets, the difference is smaller, but atmospheric shielding still remains stronger than the magnetospheric shielding.

The values of Table \ref{tab-cr-dose} have to be compared to the terrestrial
background radiation of 2.4 mSv yr$^{-1}$ \citep{Atri13}.
A planet with a sufficiently thick atmosphere (where, for example, an Earth-like N$_2$-O$_2$-atmosphere with a surface pressure of a $\sim$1 bar can be considered as ``sufficiently thick'') is protected against strong biological radiation generated by GCR regardless of its magnetic field.
For planets with a thin atmosphere, however, magnetospheric shielding is important as it can prevent an increase of the radiation dose to several hundred times the background level.
For close-in rocky planets less massive than Earth, atmospheric escape can play an important role, so that the planetary atmosphere is
likely to be less dense than on Earth.
For this reason, our
result is likely to be important for potential life on the surface of sub-Earth-mass close-in rocky exoplanets.

\section{Conclusion}\label{sec-conclusions}

Magnetic fields on most super-Earths around M-dwarf stars are likely to be weak and short-lived, 
or even non-existent.
With this in mind, the question of planetary magnetic shielding against galactic cosmic rays becomes important 
\citep[the case of stellar cosmic rays is analyzed in a separate article,][]{TabatabaVakili15}.
We use the systematic study of GCR fluxes presented in Paper I, where we found that the flux of galactic cosmic rays to the planetary atmosphere can be increased by over three orders of magnitude in the absence of a protecting magnetic field. 

With this input, we found that these energetic particles can destroy part of the atmospheric ozone and other biosignature molecules. 
However, with less than 
20\% difference in ozone column, this has little impact on remote detection of biosignature molecules.

GCRs may also change the biologically weighted UV surface flux $W$ by up to 40\%. During a stellar flare, $W$ can increase much more, and reach values a factor of 3-4 higher than on Earth, or 200-300 times the quiescent level.
Such values can be considered as non-critical. Also, this effect is not strongly dependent on the GCR flux, but we note that the surface UV flux on M-dwarf star planets is much more variable than what we know from Earth.

Finally, part of the energetic charged particles reach the planetary surface, where they contribute to a potentially harmful radiation background and increase the effective dose rate. This increase is only a factor of a few for the case of an Earth-like atmosphere.
For planets with a thin atmosphere, 
however, magnetospheric shielding is important to protect the surface.
This may also have important implications for studies of the possibility of life on the surface of sub-Earth-mass exoplanets close to their host star.

Overall, the potential absence of magnetic shielding against galactic cosmic rays has surprisingly little effect on the planet considered. Other effects are likely to dominate, unless the planet has a weak atmosphere and a strong magnetosphere. The case is different for stellar cosmic rays, which are analyzed in a companion article \citep{TabatabaVakili15}.

\begin{acknowledgements}

This study was supported by the International Space Science
Institute (ISSI) and benefited from the ISSI Team 
``Evolution of Exoplanet Atmospheres and their Characterisation''.

\end{acknowledgements}


\bibliographystyle{elsart-harv}

\end{document}